%% file: main.tex
\documentclass[12pt]{iopart}

\usepackage[multiple]{footmisc}
\usepackage{hyperref}
\expandafter\let\csname equation*\endcsname\relax
\expandafter\let\csname endequation*\endcsname\relax
\usepackage{amsmath}
\usepackage{threeparttable}    % tables with footnotes
\usepackage{dcolumn}           % decimal-aligned tabular math columns
\newcolumntype{d}{D{.}{.}{-1}}
\usepackage{nomencl}           % automatic nomenclature generation via makeindex
\makeglossary
\usepackage{graphicx}          % Package to include figures in the document
\usepackage{caption}          
\usepackage{fancyvrb}          % extended verbatim environments
\fvset{fontsize=\footnotesize,xleftmargin=2em}  
\usepackage{lettrine}          % dropped capital at beginning of paragraph
\usepackage{hyperref}          % embedding hyperlinks [must be loaded after dropping]
\graphicspath{{figs/}}         % path where all the figures are saved
\usepackage{float}             % used to place figures and tables 
\usepackage{upquote}
\usepackage{placeins}
\usepackage[retainorgcmds]{IEEEtrantools}
\usepackage{subfig}
\usepackage[export]{adjustbox}
\usepackage[utf8]{inputenc}
\usepackage[nameinlink,capitalise]{cleveref}
\usepackage[
top=2.5cm,
bottom=2.5cm,
left=2.5cm,
right=2.5cm
]{geometry}

\usepackage{siunitx}
\newcommand{\be}{\begin{equation}}
\newcommand{\ee}{\end{equation}} 
\newcommand{\eqnref}[1]{(\ref{#1})}

\usepackage{cite}
\usepackage{tikz}
\usetikzlibrary{positioning,shadows,arrows,shapes.multipart,decorations.pathmorphing}

\makeatletter
\newcommand\footnoteref[1]{\protected@xdef\@thefnmark{\ref{#1}}\@footnotemark}
\makeatother

\begin{document}

\title[]{Impact of non-equilibrium radiation in a high-enthalpy inductively coupled plasma wind tunnel}

\author{Sanjeev Kumar$^{1,*}$, Sung Min Jo$^{2,*}$, Alessandro Munaf\`{o}$^{1,3}$ and Marco Panesi$^{1,3,**}$}

\address{$^1$Center for Hypersonics and Entry Systems Studies (CHESS), \\ University of Illinois Urbana-Champaign, Urbana, IL 61801, USA}

\address{$^2$UCF Center of Excellence in Hypersonics and Space Propulsion (HYPERSPACE), \\ Department of Mechanical and Aerospace Engineering, \\ University of Central Florida, Orlando, FL 32816, USA}

\address{$^3$Department of Mechanical and Aerospace Engineering, \\ University of California Irvine, Irvine, CA 92697, USA}

\address{$^*$The two authors contributed equally to this paper.}

\address{$^{**}$Corresponding author (mpanesi@uci.edu).}

%%%%%%%%%%%%%%%%%%%%%%%%%%%%%%%%%%%%%%%%%%%%%%%%%%%%%%%%%%%%%%%%%%

%\ead{mpanesi@illinois.edu}
\vspace{10pt}
%\begin{indented}
%\item[]May 2023
%\end{indented}

\begin{abstract}
High-power inductively coupled plasma (ICP) wind tunnels are widely used to reproduce high-enthalpy environments relevant to atmospheric entry and hypersonic testing. Despite their importance, radiative heat transfer in ICP facilities is commonly neglected or modeled using simplified optically thin assumptions, and the impact of non-equilibrium radiation on plasma dynamics remains poorly quantified. In this work, a loosely coupled, multi-physics framework is developed to systematically investigate radiative cooling effects in the \SI{350}{kW} Plasmatron X facility at the University of Illinois Urbana-Champaign. The approach self-consistently couples a magnetohydrodynamic plasma framework with a spectral radiative transport solver, eliminating the need for optically thin or empirical models. Simulations are performed for nitrogen and air plasmas over a wide range of operating pressures (1–101 \si{\kilo\pascal}) and powers (100–350 \si{\kilo\watt}). The results reveal a strong pressure dependence of radiative losses, with radiation contributing negligibly at low pressures, but becoming a dominant energy sink at elevated pressures. At atmospheric pressure, radiative losses account for up to approximately 32\% and 22\% of the input power for nitrogen and air plasmas, respectively, leading to substantial reductions in core plasma temperatures. Nitrogen plasmas consistently exhibit higher radiative losses than air as a result of increased concentrations of radiatively active species and higher electron number densities. Pressure–power maps of radiative heat loss relative to input power are constructed to quantify combined operating effects and to provide guidance for facility operation and modeling fidelity. Finally, an assessment of self-absorption demonstrates that the Plasmatron X torch operates predominantly in an optically thin regime, even at the highest power and pressure conditions considered.

\end{abstract}

%
% Uncomment for keywords
\vspace{1pc}

% Uncomment for Submitted to journal title message
%\submitto{}

% Uncomment if a separate title page is required
\maketitle
 
% For two-column output uncomment the next line and choose [10pt] rather than [12pt] in the \documentclass declaration
%\ioptwocol
%

%%%%%%%%%%%%%%%%%%%%%%%%%%%%%%%%%%%%%%%%%%%%%%%%%%%%%%
\input{Intro}

%%%%%%%%%%%%%%%%%%%%%%%%%%%%%%%%%%%%%%%%%%%%%%%%%%%%%%
\input{Physical_modeling}
\input{framework}

%%%%%%%%%%%%%%%%%%%%%%%%%%%%%%%%%%%%%%%%%%%%%%%%%%%%%%
\input{simulation_setup}

%%%%%%%%%%%%%%%%%%%%%%%%%%%%%%%%%%%%%%%%%%%%%%%%%%%%%%

\input{results}

%%%%%%%%%%%%%%%%%%%%%%%%%%%%%%%%%%%%%%%%%%%%%%%%%%%%%%
\input{Conclusions}

%%%%%%%%%%%%%%%%%%%%%%%%%%%%%%%%%%%%%%%%%%%%%%%%%%%%%%
\section*{Acknowledgments}
This work has been supported under a NASA Space Technology Research Institute Award (ACCESS, grant number 80NSSC21K1117). S Kumar has been funded by the Vannevar Bush Faculty Fellowship OUSD(RE) Grant No: N00014-21-1-295 with M. Panesi as the Principal Investigator. 
%%%%%%%%%%%%%%%%%%%%%%%%%%%%%%%%%%%%%%%%%%%%%%%%%%%%%%

\section*{References}
\bibliographystyle{aiaa}
\bibliography{mybib}

\end{document}

%% file: Intro.tex
\section{\label{sec:intro}Introduction} 

High-power inductively coupled plasma (ICP) discharges are widely utilized across industry and academia to generate large plasma volumes for a wide range of possible applications. The appeal of ICPs lies in their capacity for extended testing durations and the absence of contact elements, such as electrodes, facilitating contamination-free plasma generation\cite{boulos1994thermal}. Consequently, ICPs are typically favored over arc jets for examining thermal protection systems (TPS) in atmospheric entry vehicles. Accurate modeling of plasma dynamics within ICP facilities is thus paramount for supporting experimental efforts.

Simulations of ICP discharges commonly assume Local Thermodynamic Equilibrium (LTE) conditions, where the time-scale of chemical reactions is small as compared to the flow time-scale \cite{Boulos_1976,mostaghimi1984parametric,mostaghimi1985analysis,proulx1987heating,mostaghimi1990effect,chen1991modeling,abeele2000efficient,utyuzhnikov2004simulation}. Under this assumption, all the plasma properties (\emph{e.g.,} composition, thermodynamic and transport properties) become a function of two independent state variables (\emph{e.g.,} p and T, correspondingly pressure and temperature), which drastically reduce the computational cost. To describe non-equilibrium flows, two-temperature (2T) models are widely used in the hypersonics community \cite{park1989nonequilibrium,park1993review,park2001chemical}. This model assumes fast equilibration between the heavy-species translational and the rotational energy modes, while the heavy-species electronic and vibrational energy modes are assumed to be in equilibrium with the translational energy of the free electrons. Application of the 2T model to ICP simulations can also be seen in literature \cite{mostaghimi1987two,Most_1989,atsuchi2006modeling,panesi2007analysis,zhang2016analysis,kumar2024investigation,kumar2025numerical} which reveal a significant non-Local Thermodynamic Equilibrium (NLTE) effect in the plasma and its impact on the plasma thermal and flow field. Hence, accurate modeling of non-equilibrium effects in these facilities is of utmost importance in predicting the state of the plasma. 

Physically consistent modeling of radiation presents a significant challenge in ICP simulations. Incorporating radiation requires coupling hydrodynamics-governing equations, such as Navier-Stokes, with the Radiative Transport Equation (RTE), which captures the space-time evolution of monochromatic intensity due to light emission, absorption, and scattering. This introduces substantial computational complexities since plasma optical properties are strongly dependent on wavelength, solid angle, and spatial coordinates. As a result, all the existing ICP models conveniently assume that radiative losses are optically thin \cite{boulos1976flow,miller1969temperature,owano1991nonequilibrium,abeele2000efficient,mostaghimi1984parametric}. When the plasma is assumed optically thin, radiation acts locally and may be readily taken into account via a sink term in the energy equation. The latter may be evaluated via curve-fits obtained based on experiments and/or theoretical calculations \cite{evans1967measurement,owano1990measurements,ogino2013fitting}. The use of the optically thin assumption, though suitable for continua, is rarely valid for spectral lines \cite{dresvin1972physics}. Further, these curve-fits may not be available for a variety of gases and for various operating conditions. Moreover, radiative effects in air ICPs at sub-atmospheric pressures have generally been neglected in the literature based on experimental evidence \cite{asinovsky1971experimental,devoto1978air}. However, these experiments are limited to very small diameter air plasmas (\emph{e.g.}, 2-8 mm) and necessitate careful assessment for larger plasma configurations such as the one considered in this work.

The main reason for the above simplifying assumptions being commonly used for modeling radiative effects in ICPs in the past was the computationally expensive nature of radiation solvers relying on a line-by-line method. However, recent improvements in computing power and radiative modeling efficiency have made it possible to fully couple the Navier-Stokes with RTE. Applications of the fully coupled NS-RTE approach have been widely reported in the hypersonics community to study radiative heat effects in the flow around atmospheric re-entry vehicles \cite{palmer2011direct,sahai2019flow,wood2012radiation,jo2023multi,balakrishnan1985radiative,olynick1995comparison,hartung1992development,jo2019electronic,feldick2009examination,panesi2009fire}. However, as per the authors' knowledge, there have been no reports on radiation-CFD coupling for ICP modeling. This motivates the aim of this paper, which is to study the radiative heat effects in an ICP torch using a coupled radiation-CFD approach without resorting to any simplifying assumptions such as optically thin plasma.

This work presents a systematic study of radiative heat effects in the \SI{350}{kW} Plasmatron X facility at the University of Illinois Urbana-Champaign for a broad spectrum of operating conditions for air and nitrogen mixtures. The paper is organized as follows: \cref{sec:phys_model} describes the physical model for the plasma, the electromagnetic field, and the radiation. \cref{sec:framework} describes the multi-physics coupled framework used in the work, where the main features of the three solvers are also provided. \cref{sec:problem_desc} describes the Plasmatron X facility and simulation setup. Results are presented and discussed in \cref{sec:results}. Finally, conclusions and future work are summarized in \cref{sec:conclusion}. 

%% file: Physical_modeling.tex
\section{\label{sec:phys_model}Physical Modeling}
This section presents the magnetohydrodynamic (MHD) model employed to simulate ICP discharges in this study. The model accounts for non-equilibrium plasma behavior, the self-consistent electromagnetic fields induced by the inductor coils, and the radiative heat transfer processes within the plasma domain.

\subsection{Plasma}\label{sec:plasma}
The plasmas treated in this work are modeled as a collection of neutral and charged components/species (\emph{e.g.}, $\mathrm{e}^-$, $\mathrm{N}_2$, $\mathrm{O}^+$), each behaving as an ideal gas. It is further assumed that the plasma is globally and locally neutral as well as collision-dominated, such that the use of a hydrodynamic description is appropriate \cite{boulos1994thermal}. In addition, the internal degrees of freedom for the molecular gaseous components are treated in a decoupled manner by using rigid-rotor and harmonic oscillator assumptions\cite{bottin1999thermodynamic}. Thermal and chemical non-equilibrium effects in plasma are modeled using the Park two-temperature model\cite{park1993review,park2001chemical}, where the internal states are assumed to follow a Maxwell-Boltzmann distribution at a specific temperature. Here, thermal equilibrium is assumed between the rotational and translational degrees of freedom ($T_{\mathrm{h}} = T_{\mathrm{tr}} = T_{\mathrm{r}}$) of heavy particles, and between the free electrons, electronic, and vibrational energy modes ($T_{\mathrm{e}} = T_{\mathrm{el}} = T_{\mathrm{v}}$).

Under the above assumptions, the hydrodynamics of the plasma is governed by the mass continuity, global 
momentum, total energy, and vibronic energy conservation equations \cite{Munafo_JCP_2020,Mitchner_book,gnoffo1989conservation}:
\begin{IEEEeqnarray}{lCr}
\frac{\partial \rho_{s}}{\partial t}+ \nabla  \cdot\left[\rho_{s}\left(\mathbf{v}+ \mathbf{U}_s \right)\right] = {\dot{\omega}}^{\textsc{c}}_{s} + {\dot{\omega}}^{\textsc{r}}_{s}, \quad s \in \mathcal{S}, \label{eq:cont} \\
\frac{\partial \rho \mathbf{v}}{\partial t}+ \nabla \cdot(\rho \mathbf{v} \mathbf{v} + p \mathsf{I}) =  {\nabla} \cdot \mathsf{\tau} + \mathbf{J} \times \mathbf{B}, \\
 \frac{\partial \rho \mathcal{E}}{\partial t}+ \nabla \cdot(\rho H \mathbf{v})= \nabla \cdot \left( \mathsf{\tau} \mathbf{v} - \mathbf{q}\right) - \Omega^{\textsc{r}} + \mathbf{J} \cdot \mathbf{E^{\prime}}, \\
\frac{\partial \rho e_{\mathrm{ve}}}{\partial t}+ \nabla \cdot\left(\rho e_{\mathrm{ve}} \mathbf{v} \right)  =  - \nabla \cdot \mathbf{q}_{\mathrm{ve}}  -p_{\mathrm{e}} \nabla \cdot \mathbf{v} + \Omega_{\mathrm{ve}}^{\textsc{c}} -  \Omega^{\textsc{r}} + \mathbf{J} \cdot \mathbf{E^{\prime}},\label{eq:ve_eq}
\end{IEEEeqnarray}
where $\mathcal{S}$ denotes the set of species, whereas the ve lower-script is introduced to distinguish between contributions for the sole \emph{vibronic} degrees of freedom from those referring to the plasma as a whole. The symbols in the governing equations \eqnref{eq:cont}-\eqnref{eq:ve_eq} have their usual meanings: $t$ denotes time, $\rho$ and $\mathbf{v}$ the mass density and mass-averaged velocity, respectively; $\rho_s$ and $\mathbf{U}_s$ the partial density and diffusion velocity of species $s$; $p_{\mathrm{e}}$ the pressure of free-electrons; $\mathcal{E} = e + \mathbf{v} \cdot \mathbf{v}/2$ and $H = \mathcal{E} + p/\rho$ the total energy and total enthalpy per unit-mass, where $e$ is the mixture thermal energy per unit-mass; $\mathsf{\tau}$ the stress tensor; $\mathbf{q}$ the heat flux vector; ${\dot{\omega}}^{\textsc{c}}_s$  and ${\dot{\omega}}^{\textsc{r}}_{s}$ the mass production rates due to collisional (\textsc{c}) and radiative (\textsc{r}) processes, respectively, with the corresponding energy exchange terms being $\Omega^{\textsc{c}}$ and $\Omega^{\textsc{r}}$; $\mathbf{J}$ the conduction current density; $\mathbf{E}$ and $\mathbf{B}$ the electric field and the magnetic induction, respectively; $\mathbf{E^{\prime}} = \mathbf{E} + \mathbf{v} \times \mathbf{B}$ the electric field in the hydrodynamic frame. 

More details on the calculation of $\dot{\omega}_s$, $\Omega_{\mathrm{ve}}^{\textsc{c}}$, as well as on the evaluation of thermodynamic and transport properties, can be found in \cite{kumar2024investigation,munafo2025plato}. The expression for $\Omega^{\textsc{r}}$ is detailed in \cref{sec:rad_field}.

\subsection{Electromagnetic field}\label{sec:em_field}
Electromagnetic phenomena inside ICPs are governed by Maxwell's equations \cite{Mitchner_book}. A detailed derivation of the governing equations for the electromagnetic field within ICPs can be found in \cite{kumar2024investigation}. Here, only the main expressions are summarized for the sake of completeness and brevity. 

For a two-dimensional axi-symmetric configuration, and assuming that in steady-state all variables undergo harmonic oscillations (\emph{e.g.,} $E (r, z, \, t)  =  \tilde{E} (r,z) \exp\left({\imath \omega t}\right)$), the equation governing the spatial dependence of the induced electric field consists of a scalar equation for its toroidal component:
\begin{equation}
 \dfrac{\partial}{\partial r} \left(\dfrac{1}{r} \dfrac{\partial r \tilde{E}}{\partial r}  \right) + \dfrac{\partial^2 \tilde{E}}{\partial z^2} - \imath \omega \mu_0\sigma\tilde{E} = \imath \omega \mu_0 \tilde{J_s},
\end{equation}
where $r$ and $z$ are the radial and axial coordinates, respectively. The electric field amplitude is assumed complex (\emph{i.e.}, phasor) $\tilde{E} =\tilde{E}_{\mathrm{R}} + \imath \tilde{E}_{\mathrm{I}}$, where the subscripts R and I refer to its real and imaginary components, respectively, while $\imath = \sqrt{-1}$ denotes the imaginary unit. The angular frequency of the current flowing through the inductor is $\omega = 2\pi f$. The symbol $\tilde{J_s}$ represents the current density contribution of the inductor coils. The magnetic induction phasor ($\tilde{\textbf{B}} =\tilde{\textbf{B}}_{\mathrm{R}} + \imath \tilde{\textbf{B}}_{\mathrm{I}}$) is obtained from Faraday's law, and reads:
\begin{equation}
\tilde{\mathbf{B}}_{\mathrm{R}}  =\frac{1}{\omega} \frac{\partial \tilde{E}_{\mathrm{I}}}{\partial z} \mathbf{e}_r-\frac{1}{\omega r} \frac{\partial r \tilde{E}_{\mathrm{I}}}{\partial r} \mathbf{e}_z, \label{eq:Bre_2D}
\end{equation}
\begin{equation}
\tilde{\mathbf{B}}_{\mathrm{I}}  =-\frac{1}{\omega} \frac{\partial \tilde{E}_{\mathrm{R}}}{\partial z} \mathbf{e}_r+\frac{1}{\omega r} \frac{\partial r \tilde{E}_{\mathrm{R}}}{\partial r} \mathbf{e}_z, \label{Bim_2D}
\end{equation}
where $\mathbf{e}_z$ and $\mathbf{e}_r$ denote the axial and radial unit vectors, respectively.

In general, ICP facilities operate at frequencies on the order of a few megahertz. Consequently, over one period of the driving frequency (1/f), the plasma can be considered to experience a time-averaged Lorentz force and Joule heating \cite{Most_1989}:
\begin{IEEEeqnarray}{rCl}
\left<\mathbf{J}\times \mathbf{B}\right>_z&=&-\frac{1}{2} \sigma \Re{\left[ \tilde{E} \tilde{B}^{*}_r  \right]}, \\
\left<\mathbf{J}\times \mathbf{B}\right>_r&=&\frac{1}{2} \sigma \Re{\left[ \tilde{E} \tilde{B}^{*}_z \right]}, \\
\left<\mathbf{J}\cdot \mathbf{E}\right>&=&\frac{1}{2} \sigma \tilde{E} \tilde{E}^*,
\end{IEEEeqnarray}
where $k^*$ and  $\Re{(k)}$ denote, respectively, the complex conjugate and the real part of $k$, while $ \tilde{B}_r$ and $\tilde{B}_z$ are, respectively, the radial and axial components of the magnetic induction phasor.

\subsection{Radiation}\label{sec:rad_field}
%% Sung Min %%%%%%%%%%%%
The transport of radiation within a plasma can be modeled using the radiative transport equation (RTE). This equation describes the temporal, spatial, directional, and spectral dependence of the monochromatic intensity as a result of emission, absorption, and scattering. For a two-dimensional axi-symmetric configuration and assuming steady-state and no scattering, the RTE reads: 
\begin{equation}
\begin{aligned}
& \frac{1}{r} \frac{\partial}{\partial r}\left[\mu r I_\lambda(\mathbf{r}, \, \mathbf{s})\right]+\frac{\partial}{\partial z}\left[\xi I_\lambda(\mathbf{r}, \, \mathbf{s})\right]-\frac{1}{r} \frac{\partial}{\partial \phi}\left[\delta I_\lambda(\mathbf{r}, \, \mathbf{s})\right] =-\kappa_\lambda(\mathbf{r}) I_\lambda(\mathbf{r}, \,\mathbf{s})+ \epsilon_{\lambda}\left(\mathbf{r}\right),
\label{eq:RTE}
\end{aligned}
\end{equation}

\noindent
where $\mu$, $\xi$, and $\delta$ are the direction cosines of the path $\mathbf{s}$; $\mathbf{r}$ the spatial position; $I_{\lambda}$, $\kappa_{\lambda}$, and $\epsilon_{\lambda}$ the monochromatic intensity, absorption, and emission coefficients, respectively, at the wavelength $\lambda$. The optical properties of the plasmas (\emph{e.g.}, $\epsilon_\lambda$) are evaluated via a Line-by-Line (LBL) method, followed by order-reduction using a Multi-Band Opacity Binning (MBOB) approach \cite{scoggins2013multi,jo2023multi,Jo_HMT_2020,Jo_POF_2019}. 

In the present study, the radiating species are: N, O, $\mathrm{N}_2$, $\mathrm{N}_2^+$, and $\mathrm{O}_2$. For atoms, bound-bound, bound-free, and free-free transitions are considered. In addition to the bound-free and the free-free transitions of the neutral atoms, the continuum radiation generated by the photo-detachment of atomic negative ions is also considered. This is because the amount of the photo-detachment continuum is comparable to that of the bound-free radiation under high-enthalpy air conditions \cite{Jo_HMT_2020}. In modeling atomic continuum radiation, the photo-ionization edge shift is taken into account following the work by Nicolet \cite{nicolet1970advanced}. For molecules, band systems ranging from vacuum ultraviolet (VUV) to near-infrared (IR) are taken into account, as summarized in \cref{table:rad_band_system}. 

\begin{table}
\centering
\caption{Molecular band systems of the present radiation model.}
\begin{tabular}{l l l }
\hline System & Range & Reference \\
\hline \multicolumn{3}{l}{$\mathrm{N}_2$} \\
 $\mathrm{B}^3 \Pi_g-\mathrm{A}^3 \Sigma_u^{+}$ & Visible & \cite{Laux1993} \\
 $\mathrm{C}^3 \Pi_u-\mathrm{B}^3 \Pi_g$ & UV & \cite{Laux1993} \\
 $\mathrm{c}^{\prime}{ }_4 \Sigma_u^{+}-\mathrm{X}^1 \Sigma_g^{+}$ & VUV & \cite{Hyun2009,CHAUVEAU2002503} \\
 $\mathrm{c}_3^{\prime 1} \Pi_u-\mathrm{X}^1 \Sigma_g^{+}$ & VUV & \cite{Hyun2009,CHAUVEAU2002503} \\
 $\mathrm{b}^1 \Pi_u-\mathrm{X}^1 \Sigma_g^{+}$ & VUV & \cite{Hyun2009,CHAUVEAU2002503} \\
 $\mathrm{b}^{\prime}{ }^1 \Sigma_u^{+}-\mathrm{X}^1 \Sigma_g^{+}$ & VUV & \cite{Hyun2009,CHAUVEAU2002503} \\
 $\mathrm{o}_3^1 \Pi_u-\mathrm{X}^1 \Sigma_g^{+}$ & VUV & \cite{Hyun2009,CHAUVEAU2002503} \\
\hline \multicolumn{3}{l}{$\mathrm{N}_2^{+}$} \\
 $\mathrm{B}^2 \Sigma_u^{+}-\mathrm{X}^2 \Sigma_g^{+}$ & Visible & \cite{Laux1993} \\
 $\mathrm{A}^2 \Pi_u-\mathrm{X}^2 \Sigma_g^{+}$ & Visible & \cite{Hyun2009} \\
 $\mathrm{C}^2 \Sigma_u^{+}-\mathrm{X}^2 \Sigma_g^{+}$ & UV & \cite{Gilmore1992} \\
\hline \multicolumn{3}{l}{NO} \\
 $\mathrm{B}^2 \Pi_r-\mathrm{X}^2 \Pi_r$ & UV & \cite{Laux1993} \\
 $\mathrm{A}^2 \Sigma^{+}-\mathrm{X}^2 \Pi_r$ & UV & \cite{Laux1993} \\
 $\mathrm{C}^2 \Pi_r-\mathrm{X}^2 \Pi_r$ & UV & \cite{Laux1993} \\
 $\mathrm{D}^2 \Sigma^{+}-\mathrm{X}^2 \Pi_r$ & UV & \cite{Laux1993} \\
 $\mathrm{B}^{\prime 2} \Delta-\mathrm{X}^2 \Pi_r$ & VUV-UV & \cite{Laux1993} \\
 $\mathrm{E}^2 \Sigma^{+}-\mathrm{X}^2 \Pi_r$ & VUV-UV & \cite{Laux1993} \\
\hline \multicolumn{3}{l}{$\mathrm{O}_2$} \\
 $\mathrm{B}^3 \Sigma_u^{-}-\mathrm{X}^3 \Sigma_g^{+}$ & UV & \cite{Laux1993} \\
\hline
\label{table:rad_band_system}
\end{tabular}
\end{table}

For non-Boltzmann radiation calculations, the population of excited electronic states is computed assuming a quasi-steady-state (QSS) for simplicity. This approximation transforms the electronic master equations into a set of algebraic equations. The needed electronic state-resolved rate coefficients, such as those for electron and heavy-particle impact excitation, are taken from previous studies from one of the authors \cite{jo2019electronic,Jo_HMT_2020}. As a result, the $i$-th electronic (\emph{e.g.}, emitting) state population with the non-Boltzmann effect $n_{i,\mathrm{NB}}$ is defined as:

\begin{equation}
n_{i,\mathrm{NB}} = n_{i,\mathrm{B}} \frac{\rho_i}{\chi},
\end{equation}

\noindent
where the subscripts $\mathrm{NB}$ and $\mathrm{B}$ denote the non-Boltzmann and Boltzmann state populations, respectively, while $\rho_i$ and $\chi$ are the normalized state population and the normalized species density, respectively, corresponding to the equilibrium state at a given set of flow properties.

%spectral module in the RTE solver, summarized in the previous studies of one of the co-authors \cite{jo2023multi,Jo_HMT_2020,Jo_POF_2019,Jo_PRE_2019}. \textcolor{red}{SUMMARIZE CONSIDERED RADIATION SYSTEM AS A TABLE}

In this work, the RTE (\cref{eq:RTE}) is solved using a finite-volume method for radiative transfer. The discretization is performed by defining a median-dual control volume around a nodal point on an unstructured grid topology. The intensity flux component across a control volume face is evaluated using the step scheme \cite{chai2018finite}, which corresponds to a first-order upwind approximation. To avoid control angle overlap in the non-orthogonal grid topology, a bold approximation \cite{asllanaj2007solution} is applied. The discretized system of equations is then solved using the \textsc{gmres} \cite{saad1985generalized} (General Minimal Residual) algorithm.
Once \cref{eq:RTE} is solved, the net volumetric power loss due to radiation follows from integration of the intensity over the solid angles and the wavelength:
\begin{equation}
\Omega^{\textsc{r}}=\int_{0}^{\infty} \!\!\! \kappa_{\lambda}\left[4\pi B_{\lambda} - \oint I_{\lambda} \, d\Omega\right]d\lambda,
\label{eq:RadiationLoss}
\end{equation}
\noindent
where $B_{\lambda}$ denotes the Planck function. When the plasma is considered optically thin, the second term on the right-hand side of the above equation is omitted.

%% file: framework.tex
\section{\label{sec:framework}Computational Framework}
The simulations in this study utilize a multi-solver coupled numerical framework for ICP wind tunnels, as detailed in \cite{kumar2024investigation}, which serves as the foundational platform for the present work and will be referred to as the base ICP framework. Here, only a concise overview of the latter framework is provided to contextualize the novel contributions of this work. 

The flow governing equations are solved in a block-structured finite volume (FV) solver, \textsc{hegel} (High fidElity tool for maGnEtogas-dynamic appLications) \cite{munafo2024hegel}, and the electromagnetic (EM) field is solved in a mixed finite element (FE) solver,
\textsc{flux} (Finite-element soLver for Unsteady electromagnetiX) \cite{kumar2022self}. \textsc{hegel} is coupled with the \textsc{plato} library \cite{munafo2025plato}, which takes care of the evaluation of thermodynamic and transport properties as well as mass and energy transfer source terms due to collisional chemical-kinetics processes. The plasma and the electromagnetic field are coupled through the Lorentz force ($\mathbf{J} \times \mathbf{B}$) and Joule heating ($\mathbf{J} \cdot \mathbf{E}$) terms in the fluid momentum equation and energy equations, respectively, and the electrical conductivity ($\sigma$) in the induction equation. This interaction between the plasma and the electromagnetic solver enables the accurate simulation of intricate magnetohydrodynamic phenomena within ICP wind tunnels. The coupling between \textsc{hegel} and \textsc{flux} occurs throughout the plasma domain, and both solvers require matching grids within the torch region.

\begin{figure}[h]
\begin{center}
\includegraphics[scale=0.35]{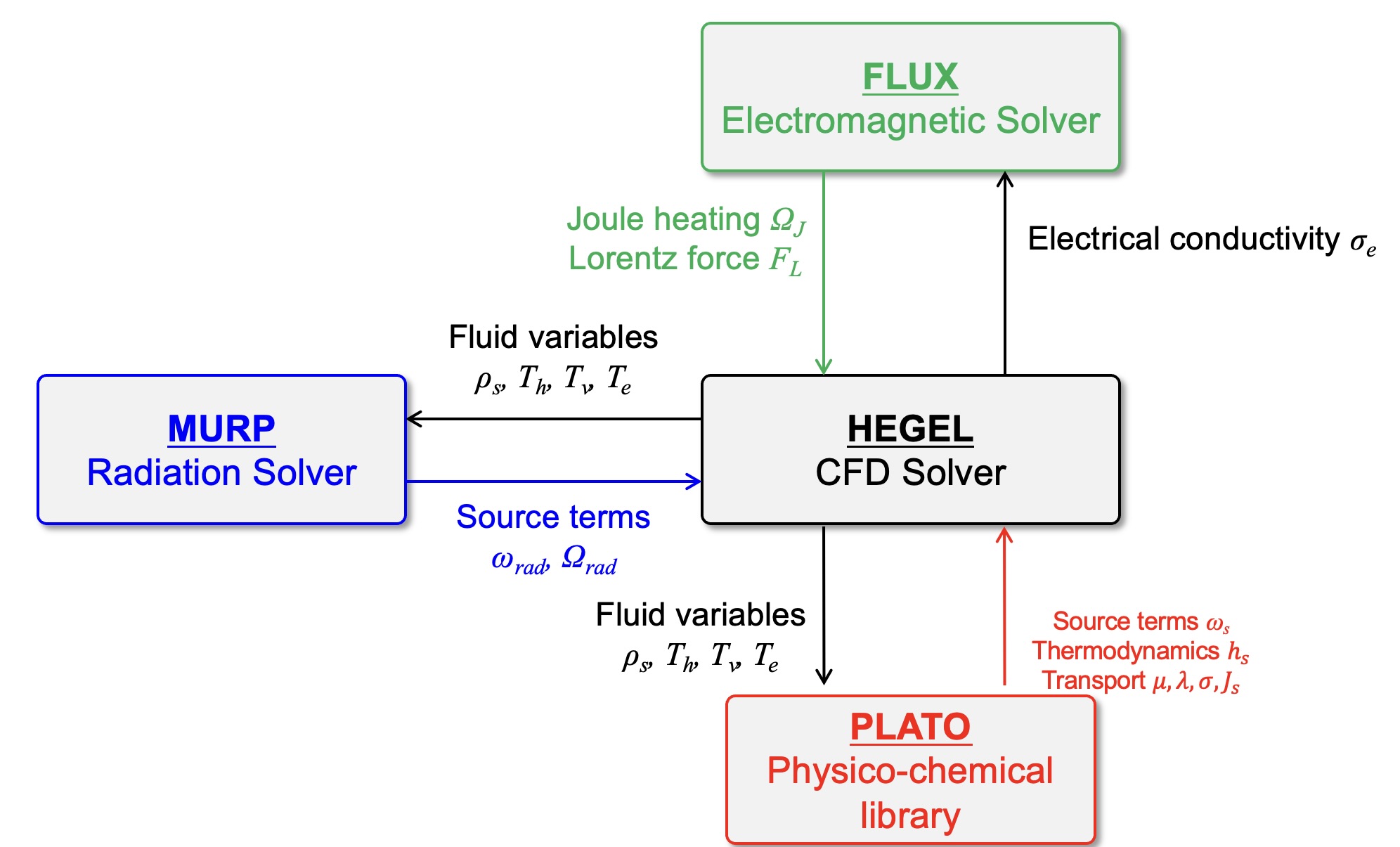}
\caption{Flowchart of the coupling framework used in this work.}
\label{fig:coupling}
\end{center}
\end{figure}

To extend the capability of the base ICP framework to study radiation transport effects, an in-house radiation solver, \textsc{murp} (MUlti-fidelity Radiation Package) \cite{jo2023multi}, has been coupled to \textsc{hegel}. \textsc{murp} is a finite volume solver, which supports integration of RTE in one-, two-, and three-dimensional spatial grid topologies. It computes the radiative heat source term $\Omega^\mathrm{R}$ for the given flow quantities, such as the plasma composition and temperatures from \textsc{hegel}, and communicates it back to \textsc{hegel}. As for the coupling with \textsc{flux}, the two solvers require matching grids in the region of overlap. \cref{fig:coupling} shows the overall coupling framework used in this work.

%% file: simulation_setup.tex
\section{\label{sec:problem_desc}Problem description and simulation setup}
\begin{figure}[!htb]
\centering
\includegraphics[scale=1]{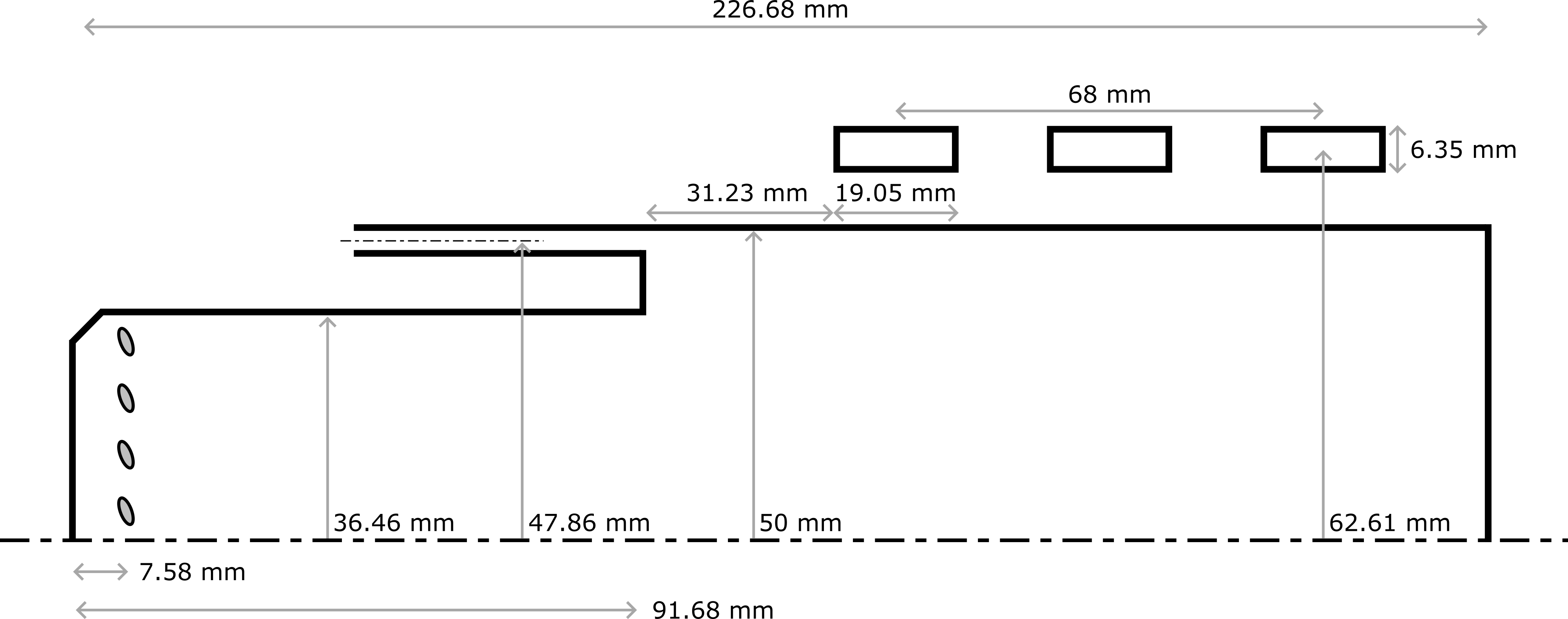}
\caption{Geometry of the Plasmatron X torch \cite{oruganti2023modeling}.}
\label{fig:pX_torch}
\end{figure}

The previously described multi-physics computational framework is applied to analyze the \SI{350}{kW} Plasmatron X facility at UIUC. \cref{fig:pX_torch} presents a schematic of the torch section \cite{oruganti2023modeling,kumar2025numerical}. The torch design features two distinct gas injection arrangements: a central injector containing 15 holes oriented downward at \SI{15}{\degree} and imparts a \SI{24}{\degree} swirl, and a sheath injector consisting of 72 axially aligned holes. In the computational model, these injectors are simplified as continuous annular inlets. For all simulations, the mass flows are maintained at \SI{0.86}{\gram/\second} for the central injector and \SI{7.13}{\gram/\second} for the sheath gas. Of the three nozzle geometries available in the Plasmatron X facility, one straight and two converging-diverging, this work considers only the straight configuration. This nozzle, placed downstream of the torch, has a length of \SI{129.54}{mm} and a diameter of \SI{100}{mm}. The induction system features a three-turn coil with a rectangular cross-section. For two-dimensional axisymmetric simulations, the coils are represented as parallel loops to preserve axial symmetry. Each coil turn is treated as an infinitesimally thin ring positioned at the inner boundary of the coil cross-section, which is a valid assumption since the skin effect causes most of the current to concentrate near this region \cite{belevitch1971lateral,blackwell2020demonstration}. The current distribution across the $N_{\mathrm{c}}$ coil turns is modeled using a point-source approximation.
    \begin{equation}
    \tilde{J_s}= J_0 \exp\left({i \omega t}\right) \sum_{i=1}^{N_{\mathrm{c}}}\delta(\mathbf{r} - \mathbf{r}_i) \, \mathbf{e}_{\theta},
    \end{equation}
    where $\mathbf{r}_i$ is the location of the center of the $i$-th coil; $\delta$ is Dirac's delta function; and $\mathrm{e}_{\theta}$ is the unit vector along the toroidal direction. All simulations are conducted at a fixed coil operating frequency of \SI{2.1}{\mega\hertz}.

It is important to note that, in actual facility operation, only a fraction of the supplied power is transferred to the plasma. This reduction arises from various losses such as imperfect electromagnetic coupling between the induction coils and the plasma, resistive losses in the coils, and inefficiencies in the matching network. Consequently, all power values presented in this study are reported together with the corresponding overall efficiency, $\eta$, defined as 
\begin{equation}
    \eta = (P - P_{\mathrm{loss}})/P
\end{equation}
where $P$ is the nominal power and $P_{\mathrm{loss}}$ the sum of all power losses. Consequently, the actual power delivered to the plasma (and which is prescribed in the simulations) is $P_{\mathrm{in}} = \eta P$.
\subsection{Domain and grids}\label{sec:mesh}
The computational domain for \textsc{hegel} and \textsc{murp} includes the torch section along with the straight nozzle. For the sake of brevity, in what follows, the torch-nozzle assembly will be simply referred to as torch.

\begin{figure}[hbt!]
\centering
\includegraphics[trim={0 5cm 0 5cm},clip,scale=0.45]{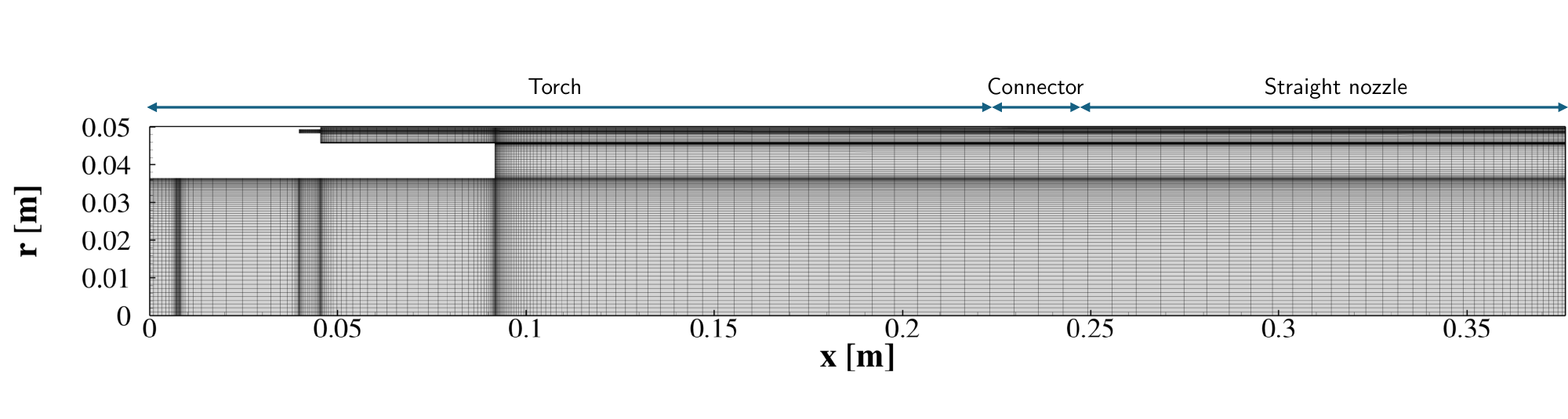}
\caption{Torch mesh (with nozzle) for the \textsc{hegel} and \textsc{flux}.}
\label{fig:pX_torch_mesh}
\end{figure}

\begin{figure}[hbt!]
\centering
\includegraphics[clip,bb=0 90 960 450,scale=0.4]{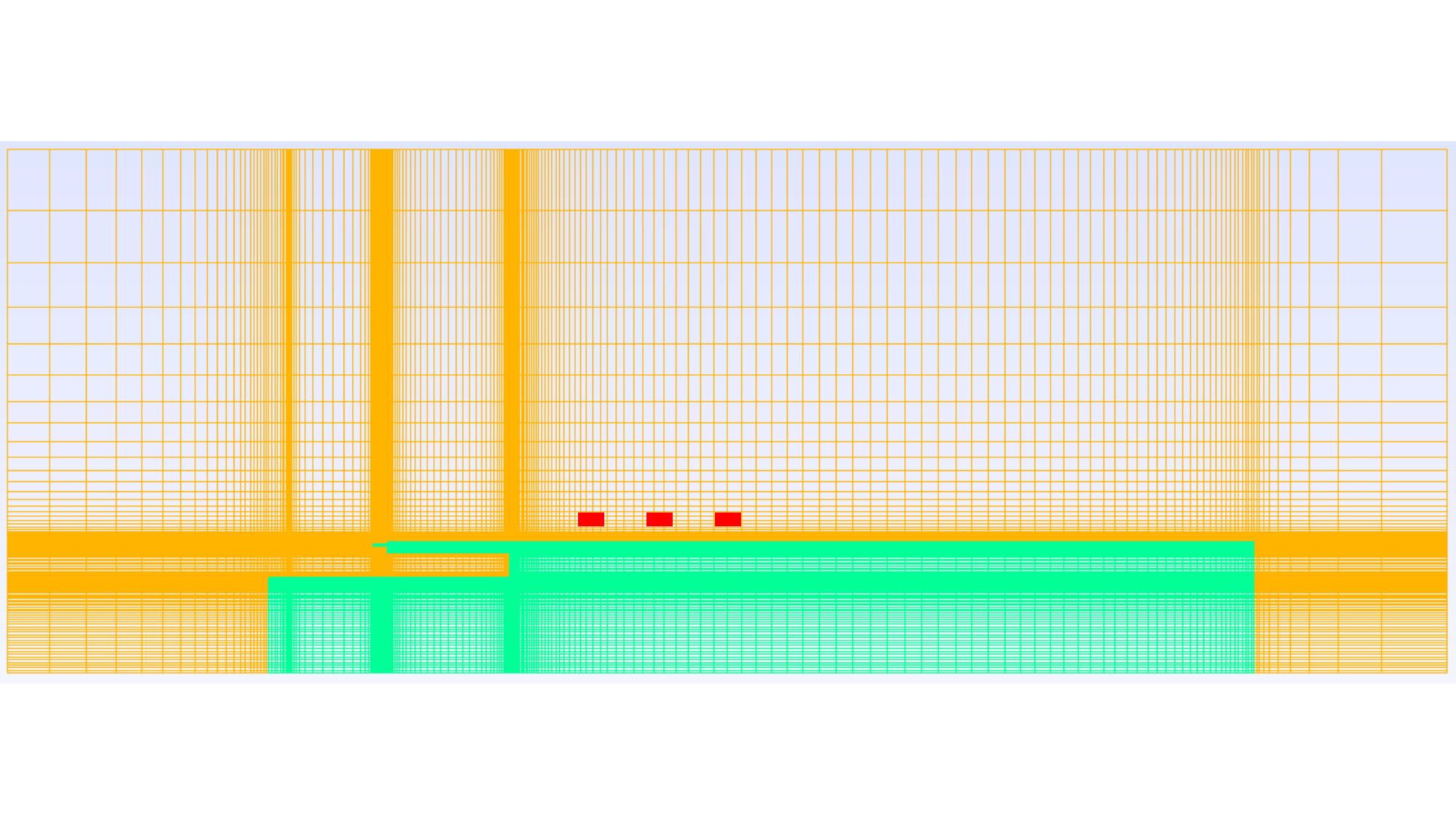}
\caption{EM solver mesh. The mesh domain highlighted in green overlaps with the plasma solver mesh. Coil locations are marked in red.}
\label{fig:pX_flux_mesh}
\end{figure}

A block-structured grid for \textsc{hegel} and \textsc{murp}, shown in \cref{fig:pX_torch_mesh}, has been used for all simulations. The grid consists of around \SI{20550}{} cells, which was found to be sufficient to provide grid-converged solutions. The computational domain for \textsc{flux} includes the torch as well as the farfield region (refer \cref{fig:pX_flux_mesh}), and consists of around \SI{41000}{} elements.

\subsection{Boundary conditions}\label{sec:BC}
\subsubsection{Plasma}
The plasma governing equations are solved by imposing the following boundary conditions:
    \begin{itemize}
    \item inlet (subsonic):
$$
\rho u=\frac{\dot{m_z}}{A}, \rho v=\frac{\dot{m_r}}{A}, \rho w=\frac{\dot{m_\theta}}{A}, \quad  y_s = y_{a,s}, \quad \frac{\partial p}{\partial x}=0, \quad T_h=T_{a} \quad \text{and} \quad T_{ve}=T_{a},
$$
where $u$, $v$, and $w$ represent the axial, radial, and swirl velocity components; $A$ the injector area; and $y_s$ the mass fraction of species $s$. The $a$ subscript denotes ambient conditions.
\item centerline (symmetry):
$$
\frac{\partial \rho_s}{\partial r}=\frac{\partial u}{\partial r}=\frac{\partial p}{\partial r}=0 \quad \text{and} \quad v=w=0.
$$
\item walls (isothermal and non-catalytic):
$$
u=v=w=0, \quad T_{\mathrm{h}}=T_{\mathrm{w}} \quad \text{and} \quad T_{\mathrm{ve}}=T_{\mathrm{w}}.
$$
\item outlet (subsonic):
$$
p=p_{\mathrm{a}} .
$$
\end{itemize} 

In all simulations, $T_{\mathrm{w}}$ and $T_{\mathrm{a}}$ are set to \SI{300}{K}.
\subsubsection{Electromagnetic field}
The electric field is zero on the axis and at large distance from the coils and the plasma (\emph{i.e.}, farfield): $E(z, 0)=0, \quad E(z, \infty)=0, \quad E( \pm \infty, r)=0$.
\subsubsection{Radiation}
The RTE is solved by imposing the following boundary conditions:
\begin{itemize}
\item walls (diffusely emitting and reflecting):
$$
I_{\lambda}\left(\mathbf{r}_{\mathrm{w}}, \, \mathbf{s}\right)=\varepsilon_{\mathrm{w}} B_{\lambda}\left(\mathbf{r}_{\mathrm{w}}\right)+\frac{1-\varepsilon_{\mathrm{w}}}{\pi} \int_{\mathbf{s}^{\prime} \cdot \mathbf{n}_{\mathrm{w}}<0} I_{\lambda}\left(\mathbf{r}_{\mathrm{w}},\, \mathbf{s}^{\prime}\right)\left|\mathbf{s}^{\prime} \cdot \mathbf{n}_{\mathrm{w}}\right| d \Omega^{\prime}\, \text { for } \mathbf{s} \cdot \mathbf{n}_{\mathrm{w}}>0,
$$
where $\mathbf{n}_{\mathrm{w}}$ is the wall outgoing unit normal. 
\item inlet and outlet (vacuum):
$$
I_\lambda \left(\mathbf{r}, \, \mathbf{s}\right) = 0 \text { for } \mathbf{s} \cdot \mathbf{n}_{\mathrm{v}}<0,
$$
where the v subscript denotes the inlet/outlet outgoing unit normal.
\end{itemize}

%A diffusely emitting and reflecting boundary condition is imposed on the wall and can be written as:

%% file: results.tex
\section{\label{sec:results}Results}
\subsection{\label{sec:general_flow}General features of the discharge}
A two-dimensional axisymmetric calculation, using a simplified geometry of the torch, has been performed by means of the Park two-temperature NLTE model to illustrate the general features of the discharge inside the Plasmatron X torch. The operating conditions are as follows: pressure \SI{5}{kPa}, power \SI{200}{kW}, and efficiency 50\%. All other parameters (\emph{e.g.}, frequency, mass flows) are fixed and retain the same values as mentioned in \cref{sec:problem_desc}. The working gas is air modeled using an eleven-species mixture: (\textbf{$\mathcal{S} = \left\{\mathrm{e}^-, \, \mathrm{N}_2, \, \mathrm{O}_2, \, \mathrm{NO},\mathrm{N},\, \mathrm{O}, \, \mathrm{N^+_2}, \, \mathrm{O^+_2},\, \mathrm{NO^+},\, \mathrm{N^+},\, \mathrm{O^+}\right\}$}). Kinetic data are taken from \cite{park2001chemical}. Radiative losses are neglected.
\begin{figure}[!htb]
%\centering
\hspace{-0.5cm}
\subfloat[][]{\includegraphics[scale=0.4]{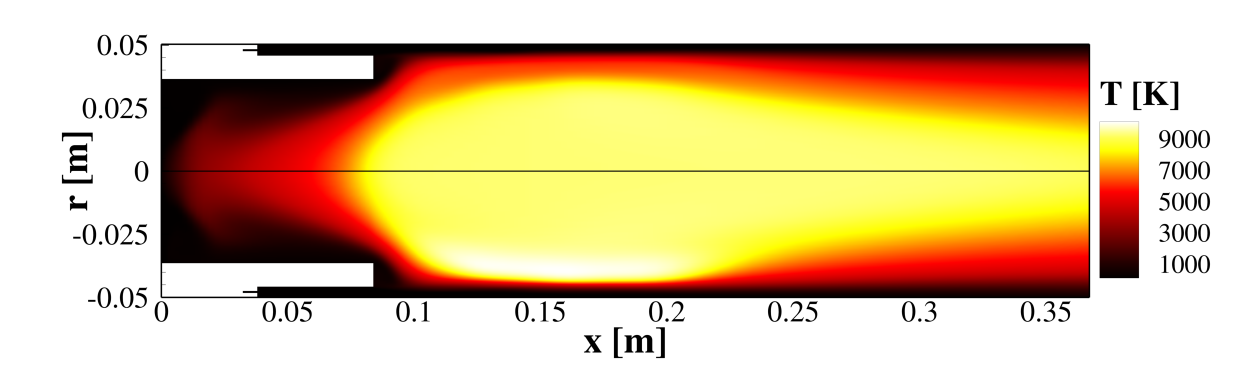}}
\subfloat[][]{\includegraphics[scale=0.4]{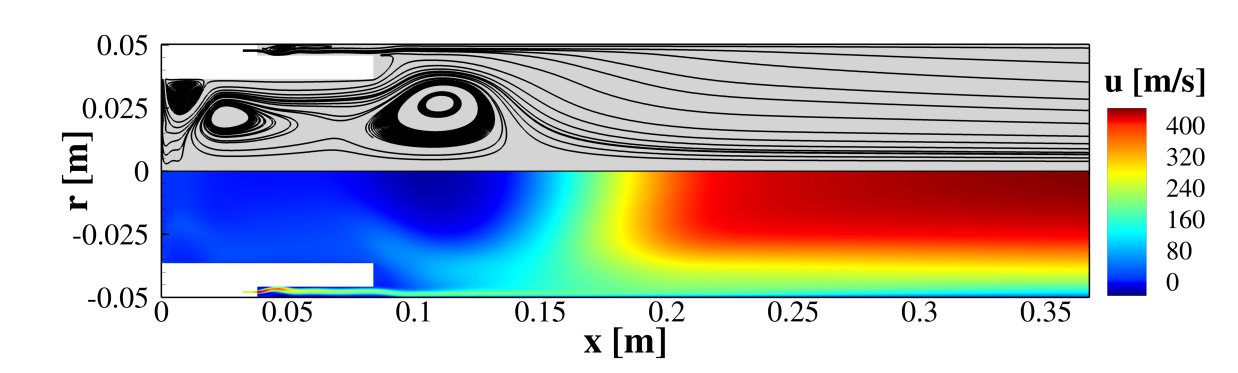}}
\caption{Plasma field inside the Plasmatron X torch. (a) Top: heavy-species temperature $\mathrm{T}_\mathrm{h}$, bottom: electro-vibrational temperature $\mathrm{T}_\mathrm{ve}$, and (b) top: streamlines, bottom: axial velocity. ($p_{\mathrm{a}} = \SI{5}{kPa}$, $P = \SI{200}{\kilo\watt}$, $\eta = 50\%$).}
\label{fig:base_flow_quantities}
\end{figure}

\begin{figure}[!htb]
%\centering
\hspace{-0.5cm}
\subfloat[][]{\includegraphics[scale=0.25]{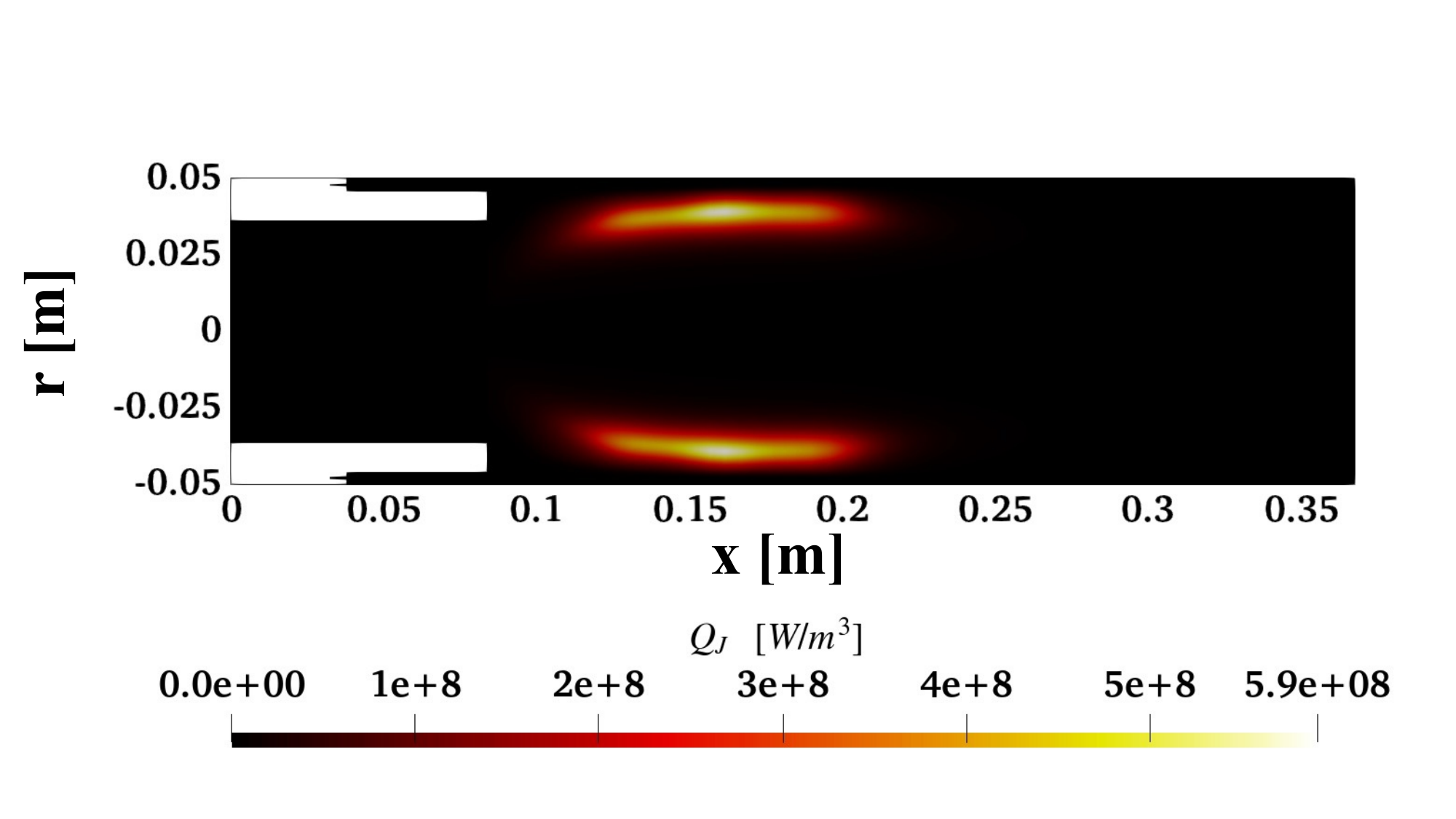}}
\subfloat[][]{\includegraphics[scale=0.25]{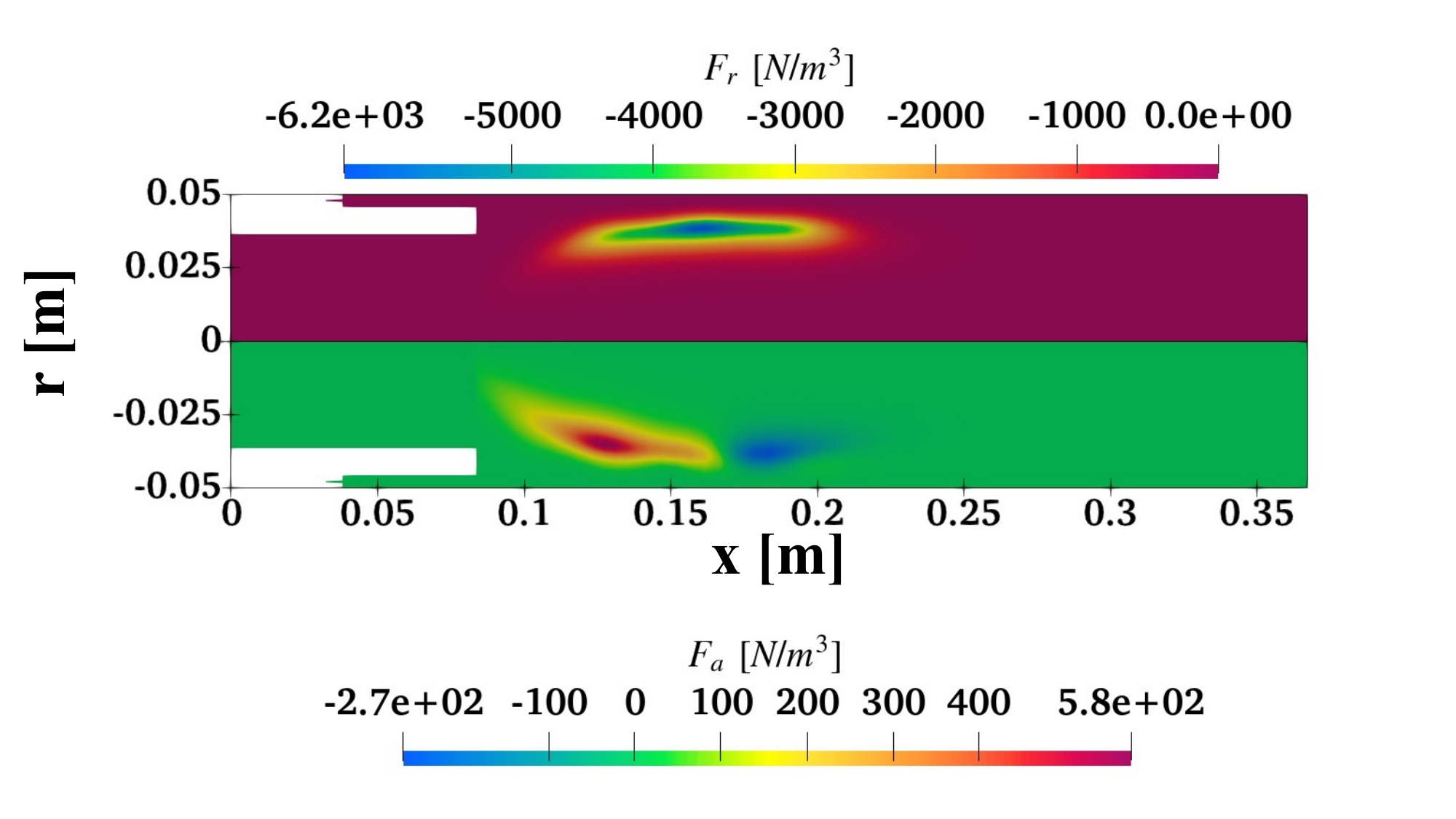}}
\caption{Distribution of the electromagnetic quantities inside the torch. (a) Joule heating, (b) Lorentz forces (top: radial component, bottom: axial component). ($p_{\mathrm{a}} = \SI{5}{kPa}$, $P = \SI{200}{\kilo\watt}$, $\eta = 50\%$).}
\label{fig:base_EM_quantities}
\end{figure}

\begin{figure}[!htb]
\centering
\subfloat[][]{\includegraphics[scale=0.5]{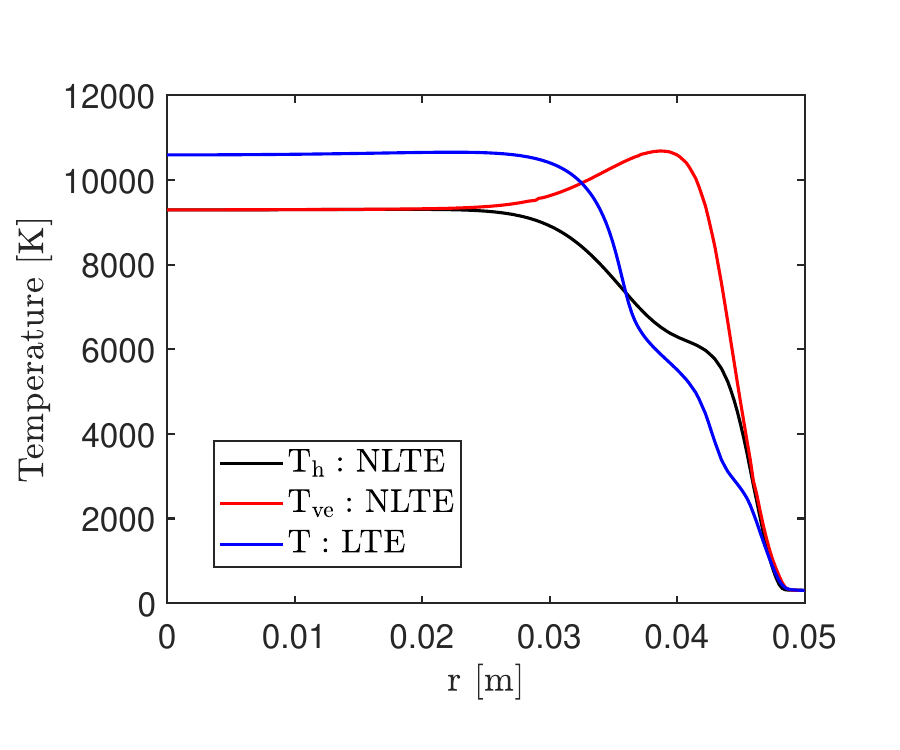}}
\subfloat[][]{\includegraphics[scale=0.5]{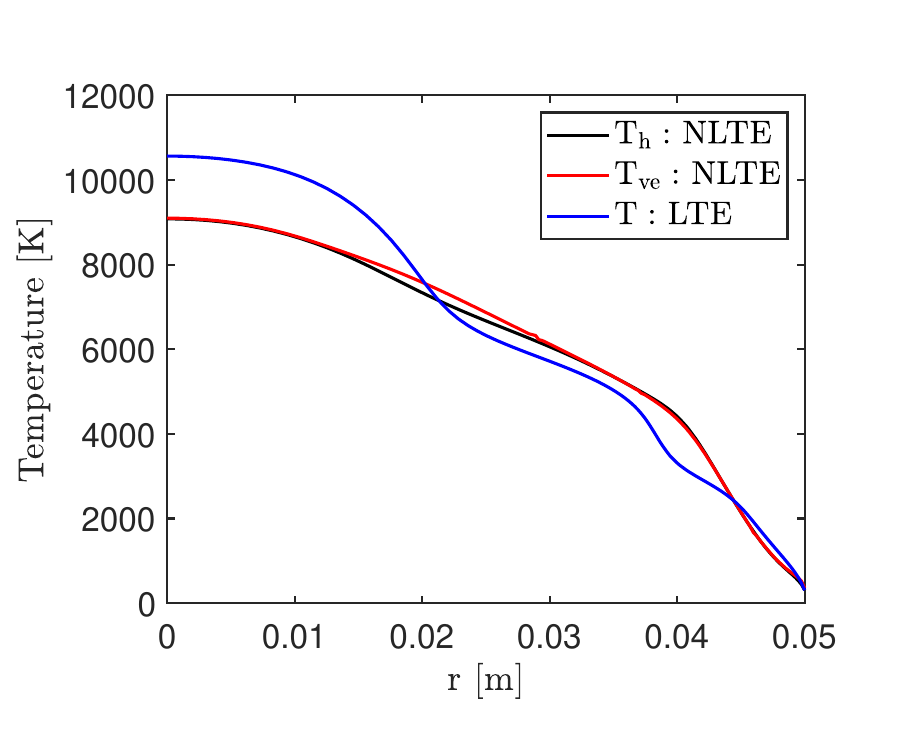}}
\caption{Radial temperature profiles obtained using the NLTE and LTE simulations. (a) x = \SI{0.15}{m} (mid-torch location), and (b) torch outlet. ($p_{\mathrm{a}} = \SI{5}{kPa}$, $P = \SI{200}{\kilo\watt}$, $\eta = 50\%$). }
\label{fig:base_temp_profiles}
\end{figure}

\begin{figure}[!htb]
\centering
\subfloat[][]{\includegraphics[scale=0.5]{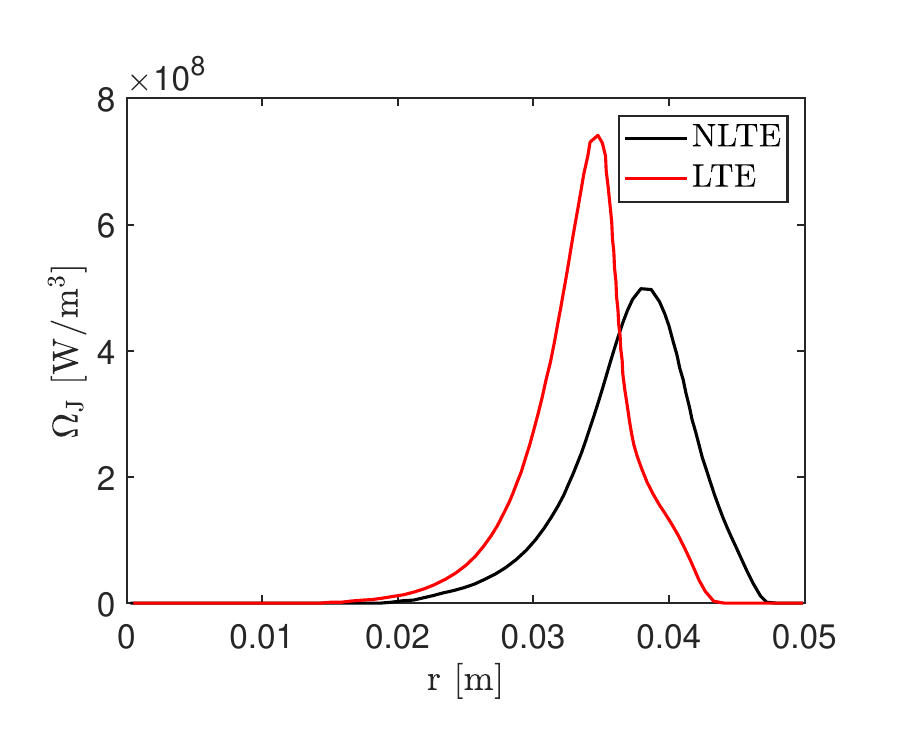}}
\subfloat[][]{\includegraphics[scale=0.5]{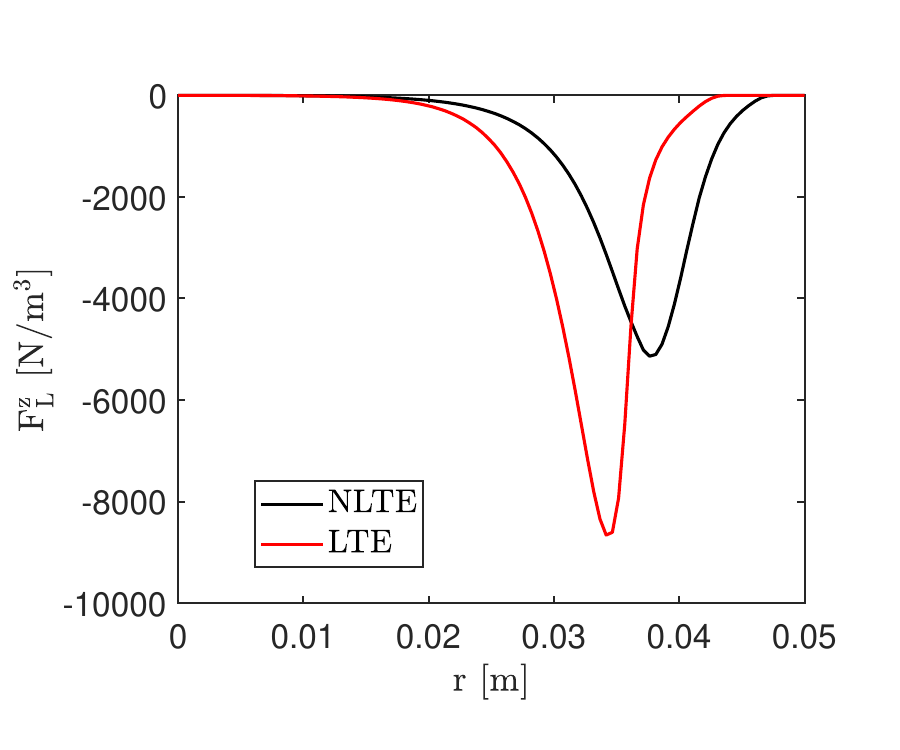}}
\caption{LTE and NLTE radial profiles of Joule heating and Lorentz force at $x = \SI{0.15}{m}$ (coil region). ($p_{\mathrm{a}} = \SI{5}{kPa}$, $P = \SI{200}{\kilo\watt}$, $\eta = 50\%$). }
\label{fig:base_EM_profiles}
\end{figure}

\cref{fig:base_flow_quantities} shows the plasma field (axial velocity and temperature contours) inside the torch. The streamlines highlight a large recirculation eddy close to the inlet, representing a typical vortex-mode discharge. As reported in the literature \cite{boulos1985inductively,eckert1974induction,abeele2000efficient}, this phenomenon plays a crucial role in maintaining the discharge. The recirculation bubble carries the hot plasma from the inductor zone back towards the inlet, preventing the plasma from getting swept away by the flow. The plasma is suspended in the middle of the inductor zone, shielded from the tube by a layer of cold fluid coming from the sheath gas injector. \cref{fig:base_EM_quantities} illustrates the distribution of Joule heating and Lorentz forces, predominantly concentrated in the coil region. Joule heating in this region raises the temperature of the incoming cold gas, forming a hot plasma core. Additionally, as shown in \cref{fig:base_EM_quantities} (b), the radial Lorentz force is directed towards the axis (as indicated by the negative sign of the values), ensuring that the hot plasma is pushed towards the axis. Meanwhile, the axial component of the Lorentz force, acting in the negative $x$-direction around $x = \SI{0.2}{m}$, draws the plasma back towards the inlet. The interplay between the swirling injector and Lorentz forces ultimately generates the large recirculation bubble, stabilizing the plasma discharge. 

The temperature field highlights significant non-equilibrium between the translational and the electro-vibrational modes in the coil region, as shown in \cref{fig:base_temp_profiles}. At low pressures, such as those considered here, the low electron-heavy collisional frequency is insufficient to establish and maintain thermal equilibrium, especially in the coil region, where the energy deposition is maximum. Away from the coils, thermal equilibrium between the translational and the electro-vibrational modes is observed. However, the effect of NLTE plasma formation propagates towards the outlet. As a result, the NLTE temperature distributions differ significantly from those obtained under the LTE assumption. This observation is in line with the findings reported in refs.\cite{zhang2016analysis,david_thesis}, and highlights the importance of using a correct physico-chemical model depending on the operating conditions. \cref{fig:base_EM_profiles} further investigates the Joule heating and Lorentz Force distributions inside the torch. The Joule heating profiles at the mid-torch location reveal that the heating profile is thicker in the case of the NLTE simulation. Hence, a larger volume of plasma is heated by the inductor for the same power, causing the peak temperature to drop significantly from the LTE simulation. Moreover, the location of the maximum departure between $T_{\mathrm{h}}$ and $T_{\mathrm{ve}}$ occurs at the location of the Joule heating peak (\emph{i.e.}, r = \SI{0.04}{m}), which further confirms the role of inductor coils in creating non-equilibrium plasma.
%%%%%%%%%%%%%%%%%%%%%%%%%%%%%%%%%%%%%%%%%%%%%%%%%%%%%%
\subsection{\label{Radiation}Effect of radiation coupling}
Having discussed the general features of the baseline discharge, we now turn to the main focus of this work, which is the study of the influence of radiative cooling.

\subsubsection{Simulation setup for radiation-coupled simulations}
Radiation-coupled simulations were performed for two mixtures: nitrogen ($\mathrm{e}^-,\mathrm{N}_2,\mathrm{N},{\mathrm{N}_2}^{+},\mathrm{N}^+$) and air ($\mathrm{e}^-,\mathrm{N}_2,\mathrm{O}_2,\mathrm{NO},\mathrm{N},\mathrm{O},\mathrm{N}_2^+,\mathrm{O}_2^+,\mathrm{NO}^+,\mathrm{N}^+,\mathrm{O}^+$). Since the impact of radiation losses strongly depend on pressure, due to its dependence on the number densities of radiating species \cite{wilbers1991radiative,ogino2013fitting,evans1967measurement,owano1990measurements}, simulations were first conducted at different pressures: \SI{1}{kPa}, \SI{5}{kPa}, \SI{10}{kPa}, \SI{30}{kPa}, and \SI{101}{kPa}, while keeping the operating power fixed at \SI{200}{kW}, to analyze the impact of radiation cooling at different pressures. The efficiency for all cases has been set to \SI{50}{\percent}. All other operating conditions retain the same values as in \cref{sec:problem_desc}. Data exchange between \textsc{hegel} and \textsc{murp} was performed every 10000 \textsc{hegel} time-steps (\emph{i.e.}, iterations), and 10-12 couplings were found to be sufficient to obtain a converged solution in all the cases. 
 
The wavelength domain, ranging from \SI{200}{\nano\meter} to \SI{1000}{\nano\meter}, was evenly subdivided (in the frequency domain) using \num{600001} spectral points at every spatial location, while the 4$\pi$-sr of the solid angle was discretized into 16$\times$32 polar and azimuthal angles, respectively. Once the plasma optical properties were evaluated based on the LBL approach (with non-Boltzmann correction, if needed), model reduction was carried out using the MBOB algorithm. The MBOB method divides the entire spectral range into a given number of bands ($\mathrm{N}_{\mathrm{BAND}}$), and then the spectral points
within a given band are further decomposed into a given number of bins ($\mathrm{N}_{\mathrm{BIN}}$). As a result, the RTE is solved $\mathrm{N}_{\mathrm{BAND}} \times \mathrm{N}_{\mathrm{BIN}}$
 times for a given control volume and a given direction. This approach leads to good performance for molecular bands. For atomic line radiation, the accuracy could be improved by varying the way the band-bin structures are constructed \cite{sahai2020comparative}. For the conditions adopted in this work, a convergence study revealed that 20 bands and 20 bins essentially led to the same results as those obtained using the LBL method for quantities of interest such as temperature and chemical composition. 
\subsubsection{Nature of radiative heat loss distribution in the torch}
\begin{figure}[!htb]
%\centering
\hspace*{-0.5cm}
\subfloat[$\Omega$\textsubscript{J} \textsubscript{peak} = \SI{680}{MW/m^3}, $\Omega$\textsuperscript{R} \textsubscript{peak} = \SI{8.1}{MW/m^3}]{\includegraphics[scale=0.18, clip, trim=5in 0.1in 5in 10in]{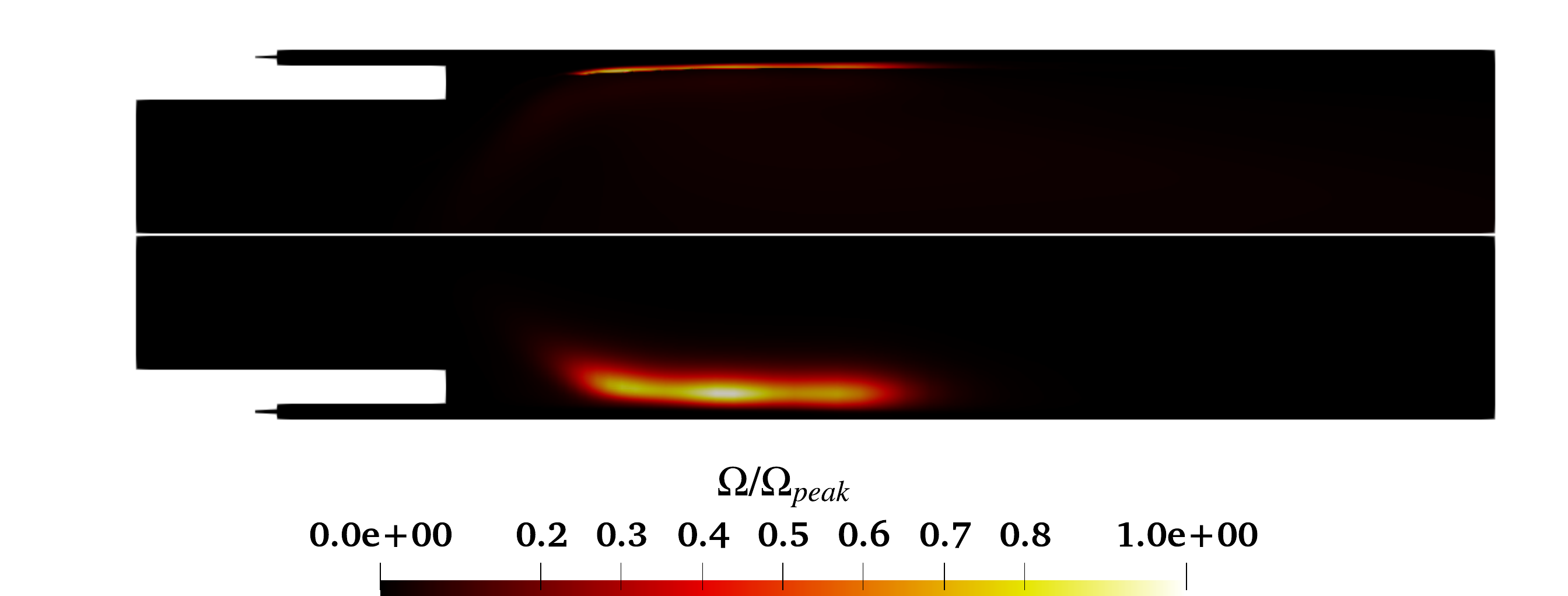}}
\subfloat[$\Omega$\textsubscript{J} \textsubscript{peak} = \SI{810}{MW/m^3}, $\Omega$\textsuperscript{R} \textsubscript{peak} = \SI{5.2}{MW/m^3}]{\includegraphics[scale=0.18, clip, trim=5in 0.1in 5in 10in]{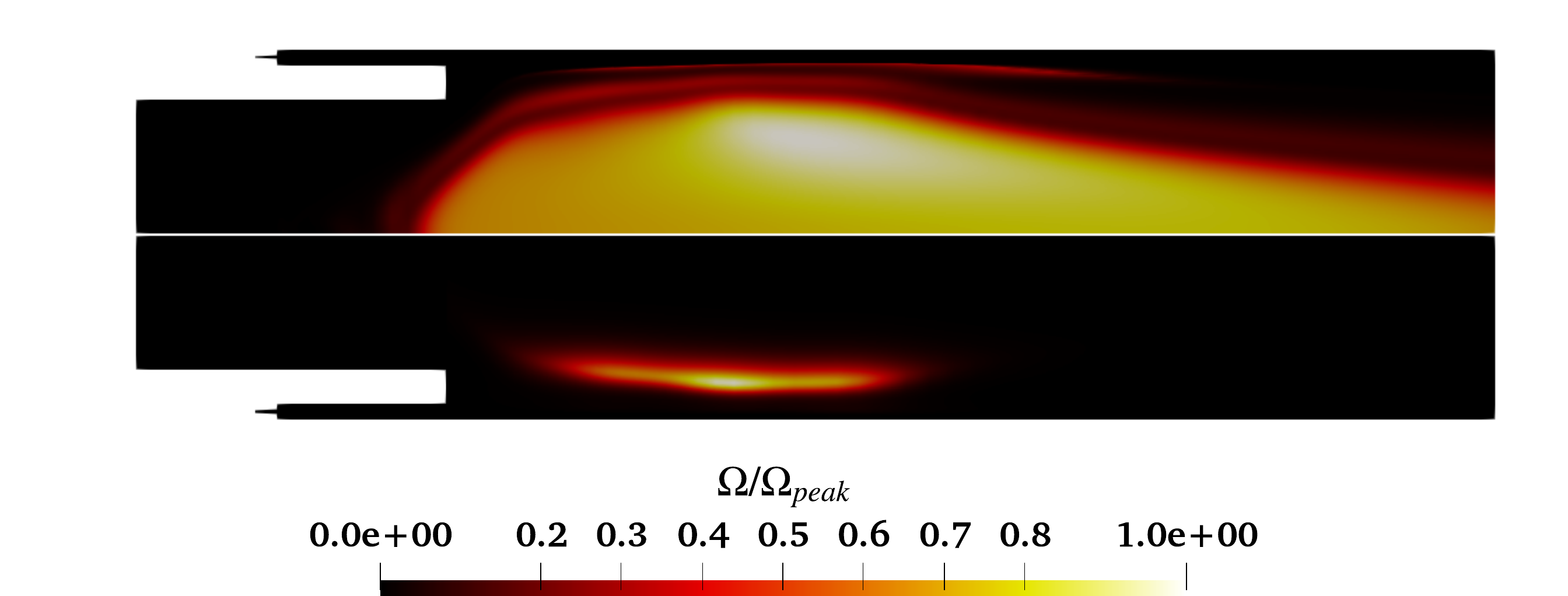}}
\\
%\centering
\hspace*{-0.5cm}
\subfloat[$\Omega$\textsubscript{J} \textsubscript{peak} = \SI{840}{MW/m^3}, $\Omega$\textsuperscript{R} \textsubscript{peak} = \SI{17.1}{MW/m^3}]{\includegraphics[scale=0.18, clip, trim=5in 0.1in 5in 10in]{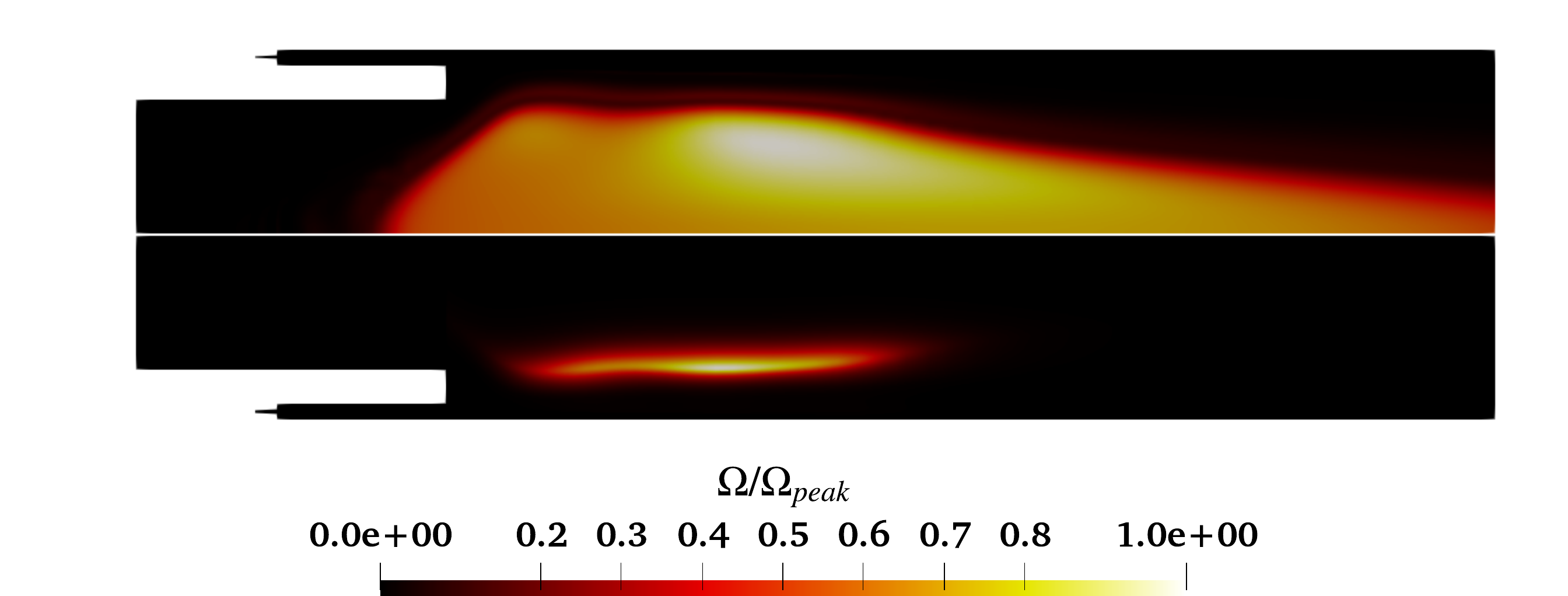}}
\subfloat[$\Omega$\textsubscript{J} \textsubscript{peak} = \SI{560}{MW/m^3}, $\Omega$\textsuperscript{R} \textsubscript{peak} = \SI{44}{MW/m^3}]{\includegraphics[scale=0.18, clip, trim=5in 0.1in 5in 10in]{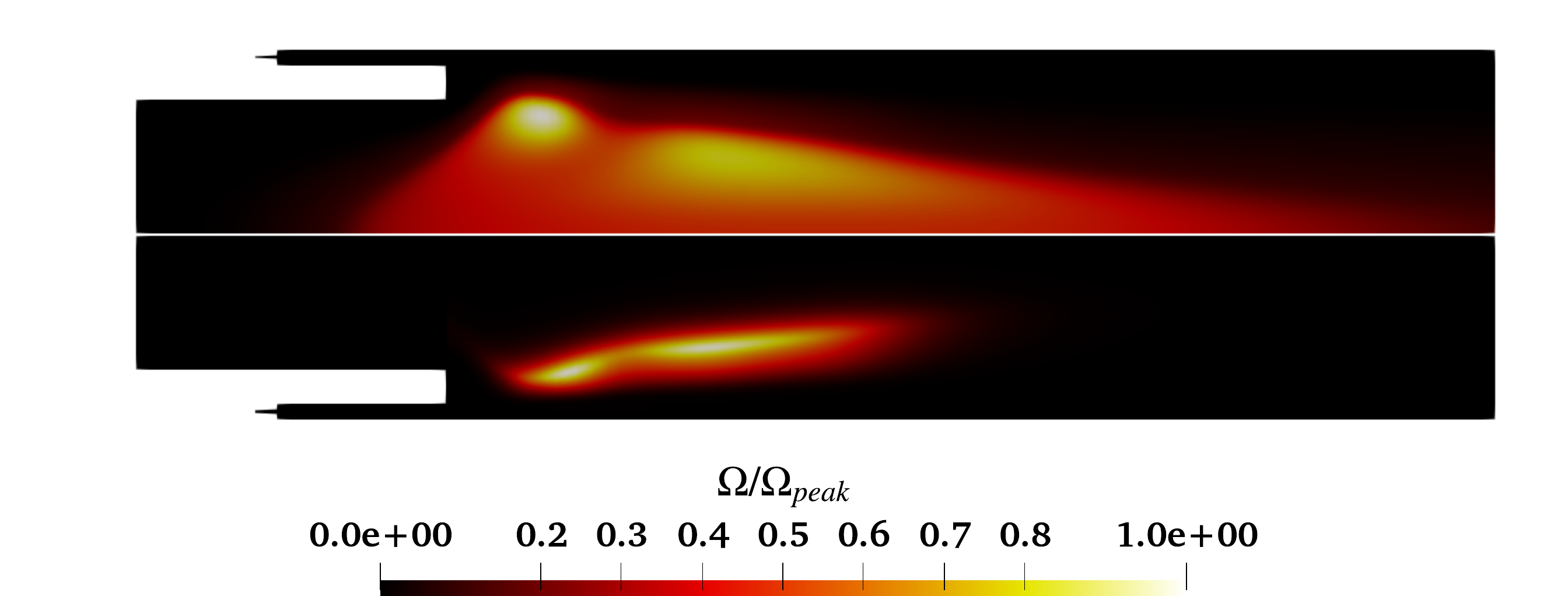}}
\caption{Radiative heat loss and Joule heating distribution normalized by corresponding peak values for nitrogen plasma: (a) 1 \si{\kilo\pascal}, (b) 5 \si{\kilo\pascal}, (c) 10 \si{\kilo\pascal} and (d) 30 \si{\kilo\pascal}, all at a fixed operating power of \SI{200}{kW} (with $\eta = 50\%$). Top: $\Omega$\textsuperscript{R}, bottom: $\Omega$\textsubscript{J}.} 
\label{fig:Qrad_N2}
\end{figure}

\begin{figure}[!htb]
\hspace*{-0.5cm}
\subfloat[$\Omega$\textsubscript{J} \textsubscript{peak} = \SI{680}{MW/m^3}, $\Omega$\textsuperscript{R} \textsubscript{peak} = \SI{7.4}{MW/m^3}]{\includegraphics[scale=0.18, clip, trim=5in 0.1in 5in 10in]{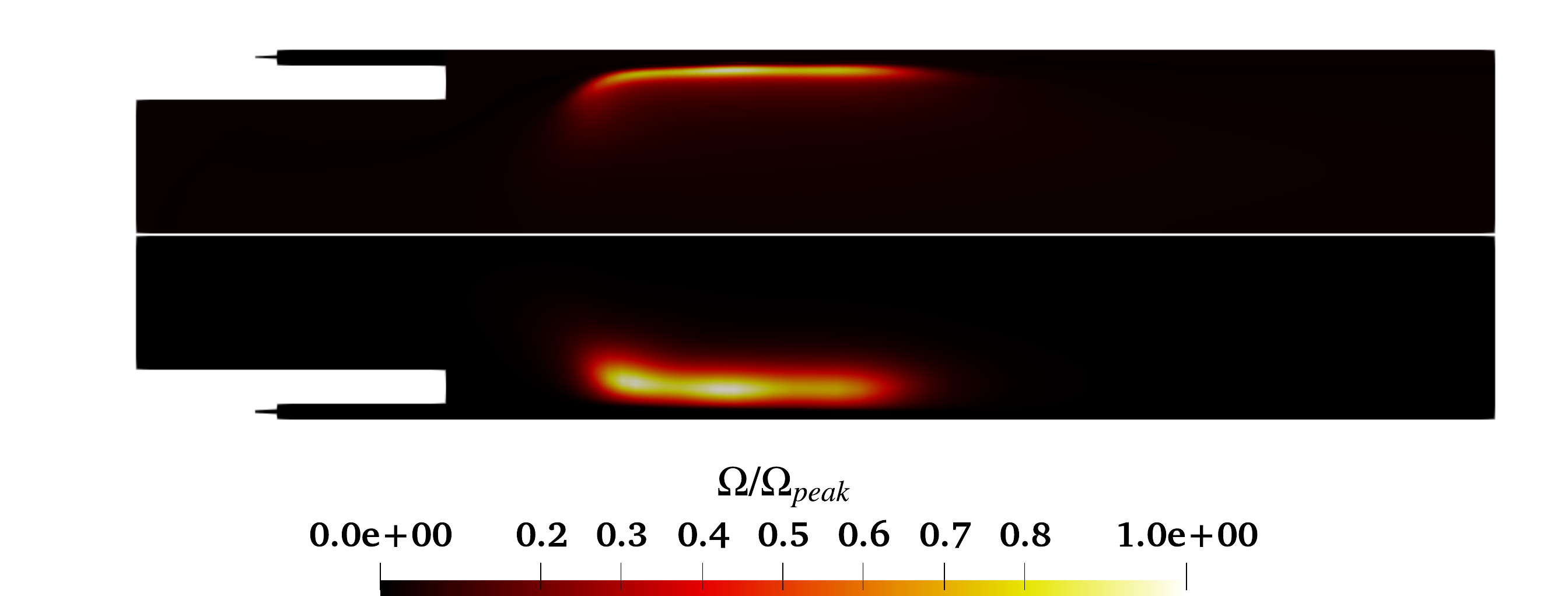}}
\subfloat[$\Omega$\textsubscript{J} \textsubscript{peak} = \SI{810}{MW/m^3}, $\Omega$\textsuperscript{R} \textsubscript{peak} = \SI{2.1}{MW/m^3}]{\includegraphics[scale=0.18, clip, trim=5in 0.1in 5in 10in]{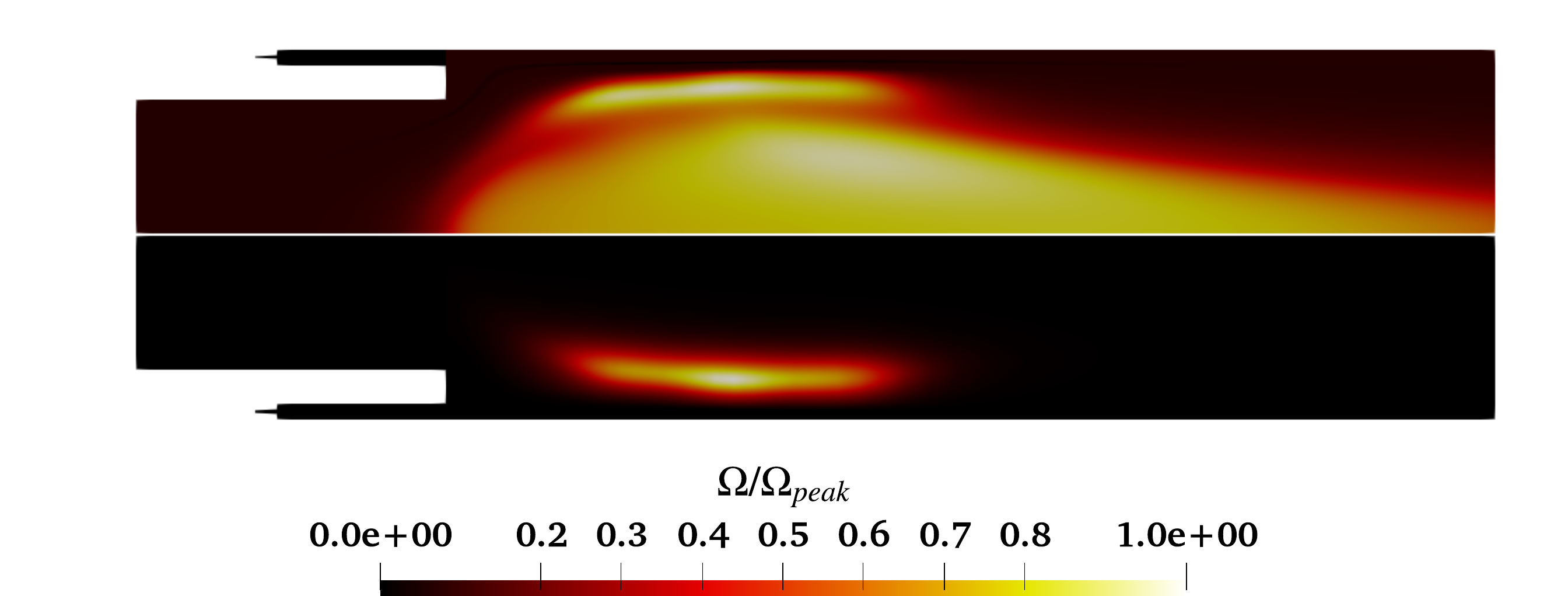}}
\\

\hspace*{-0.5cm}
\subfloat[$\Omega$\textsubscript{J} \textsubscript{peak} = \SI{840}{MW/m^3}, $\Omega$\textsuperscript{R} \textsubscript{peak} = \SI{11.5}{MW/m^3}]{\includegraphics[scale=0.18, clip, trim=5in 0.1in 5in 10in]{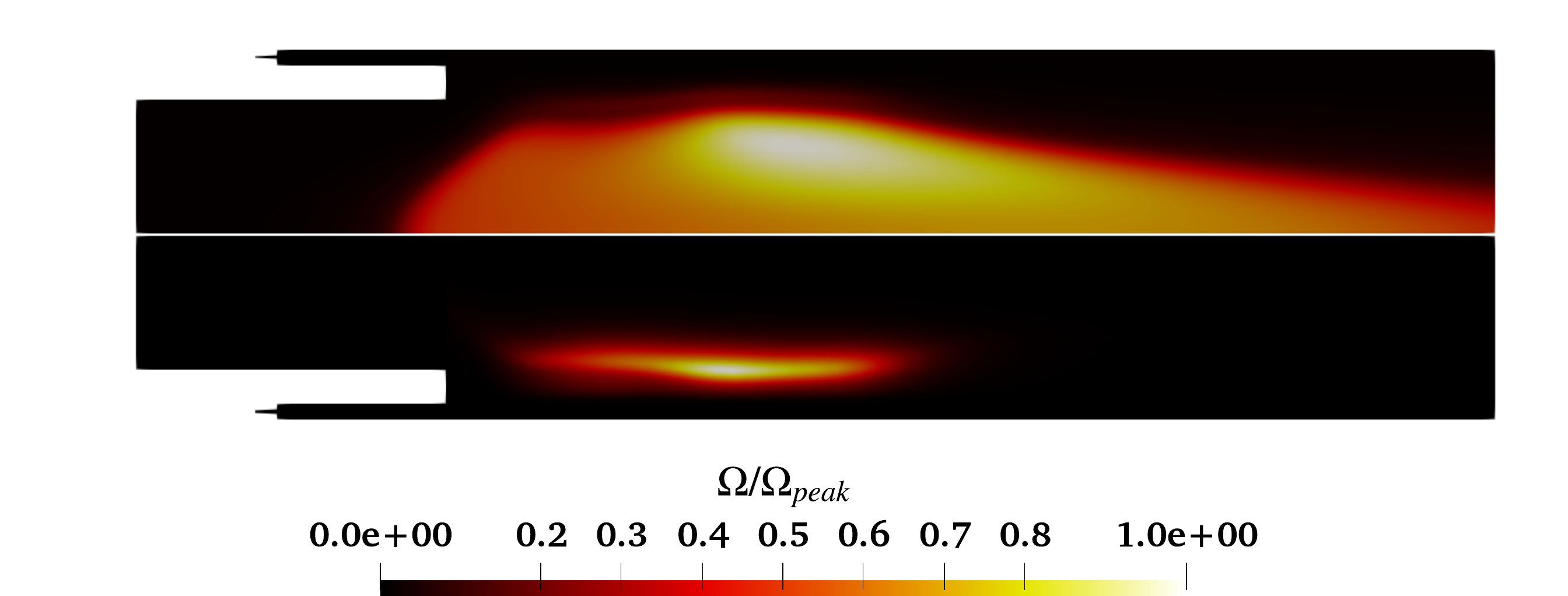}}
\subfloat[$\Omega$\textsubscript{J} \textsubscript{peak} = \SI{560}{MW/m^3}, $\Omega$\textsuperscript{R} \textsubscript{peak} = \SI{32.7}{MW/m^3}]{\includegraphics[scale=0.18, clip, trim=5in 0.1in 5in 10in]{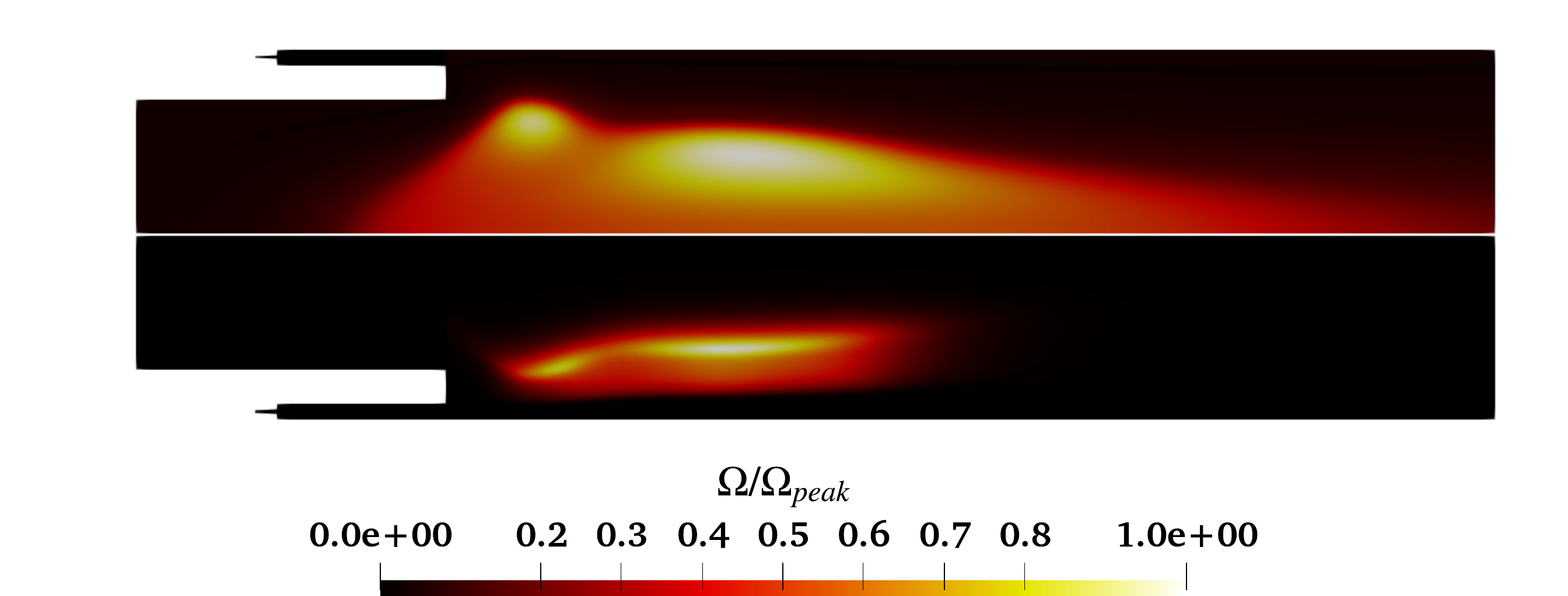}}
\caption{Radiative heat loss and Joule heating distribution normalized by corresponding peak values for air plasma: (a) 1 \si{\kilo\pascal}, (b) 5 \si{\kilo\pascal}, (c) 10 \si{\kilo\pascal} and (d) 30 \si{\kilo\pascal}, all at a fixed operating power of \SI{200}{kW} (with $\eta = 50\%$). Top: $\Omega$\textsuperscript{R}, bottom: $\Omega$\textsubscript{J}.} 
\label{fig:Qrad_air}
\end{figure}

\begin{figure}[!htb]
\centering
\subfloat[][]{\includegraphics[scale=0.475,clip, trim=1in 0in 0.5in 4in]{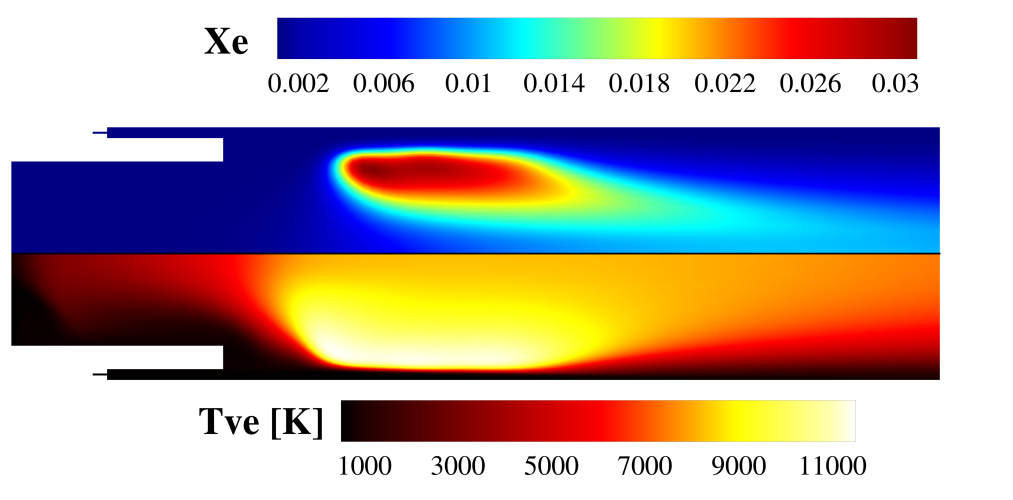}}
\subfloat[][]{\includegraphics[scale=0.475,clip, trim=1in 0in 0.5in 4in]{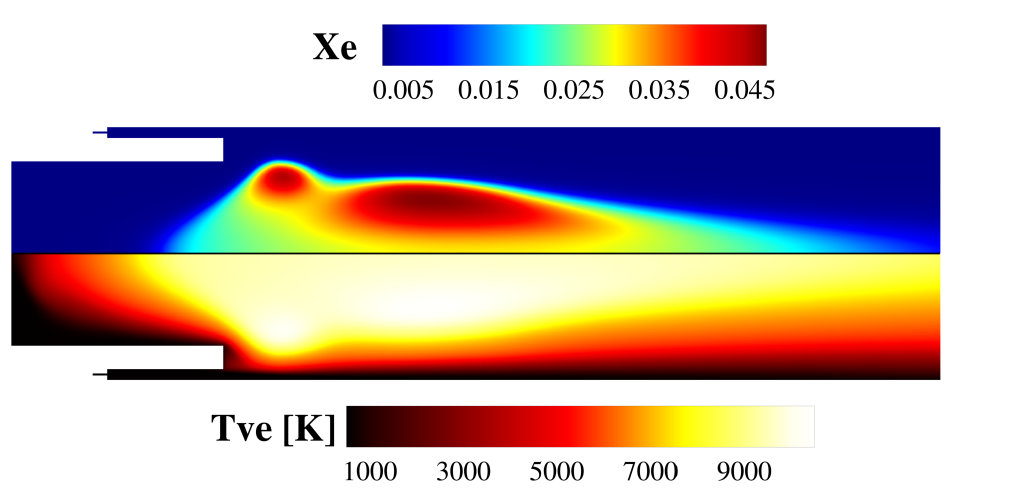}}
\caption{Distribution of electron mole-fraction $X_e$ and electro-vibrational temperature $T_{ve}$ in the torch for air plasma. (a) \SI{1}{kPa} case, and (b) \SI{30}{kPa} case, both at a fixed operating power of \SI{200}{kW} (with $\eta = 50\%$).}
\label{fig:Xe_Tve_air}
\end{figure}

\cref{fig:Qrad_N2,fig:Qrad_air} present the distributions of Joule heating and radiative heat loss within the torch at different operating pressures for nitrogen and air plasmas, respectively. The Joule heating distribution is primarily governed by the spatial distribution of the electric field and electrical conductivity, and is consistently concentrated in the coil region where the electric field intensity is maximum. In contrast, the radiative heat loss distribution exhibits a marked dependence on pressure, as evidenced by the contour plots. The variation in radiative loss distribution with pressure can be explained by \cref{fig:Xe_Tve_air}, which displays the distribution of the electron concentration and electro-vibrational temperature for air at \SI{1}{kPa} and \SI{30}{kPa}. It can be clearly seen that the radiative cooling distributions at the two pressures strongly follow the $X_{\mathrm{e}}$ and $T_{\mathrm{ve}}$ distributions, a trend also observed for nitrogen. This observation agrees with the theoretical results in ref. \cite{ogino2013fitting}, where the optically thin radiative loss term in air plasmas was found to be strongly dependent on electron temperature and number density. Nevertheless, an assessment of whether the plasma in the Plasmatron X torch operates in an optically thin or thick regime is presented in \cref{sec:optical_thickness}, following the definition of the relevant characterization parameters.

\subsubsection{Quantification of the total radiative power loss}
\begin{figure}[h]
\begin{center}
\includegraphics[scale=0.4]{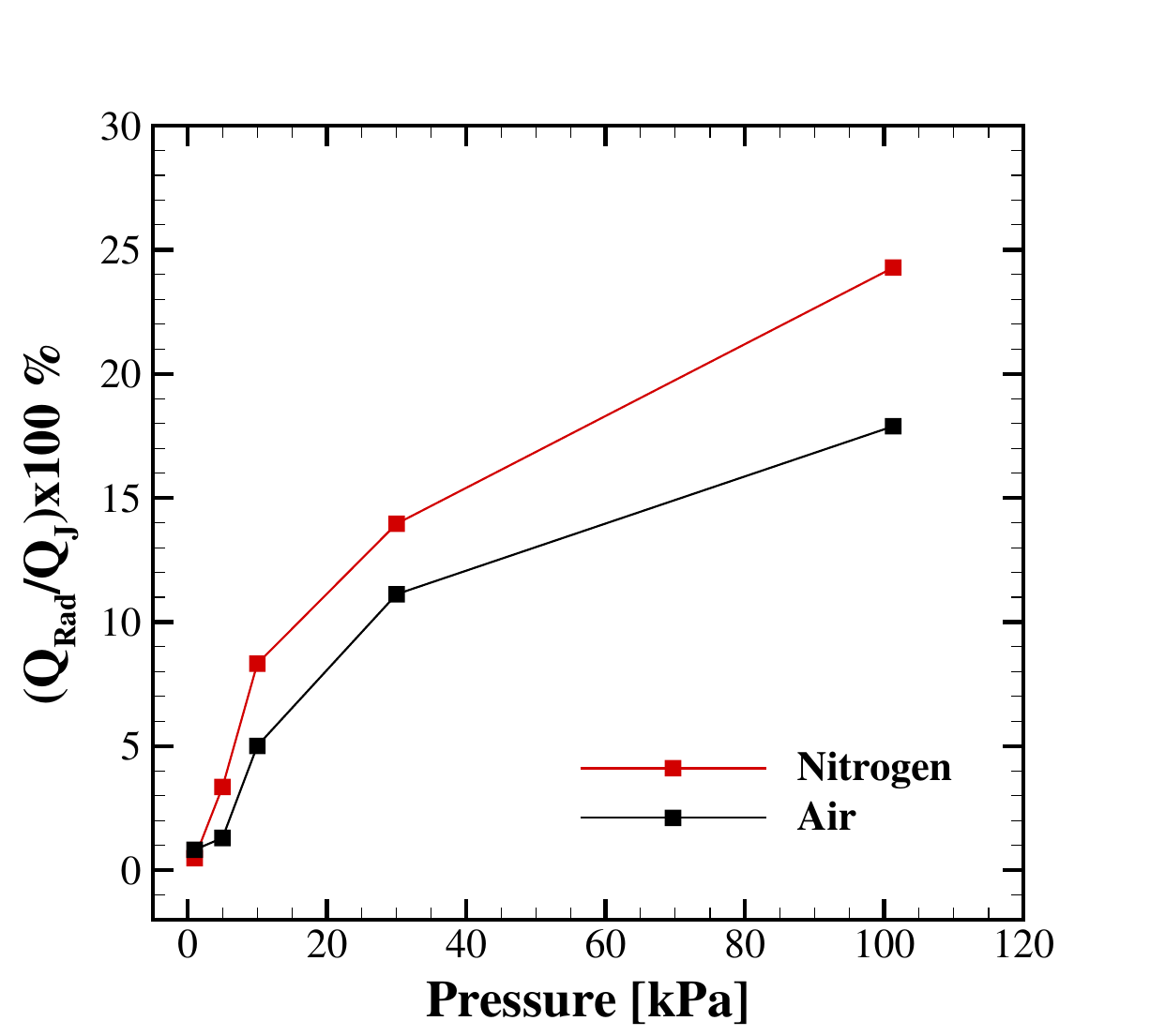}
\caption{Total radiative heat loss in the ICP torch as a percent of input power at a fixed operating power of \SI{200}{kW} (with $\eta = 50\%$). }
\label{fig:Qrad_vs_pr}
\end{center}
\end{figure}

\begin{figure}[hbt!]
\centering
\includegraphics[scale=0.6,clip, trim=1in 0in 0.5in 4in]{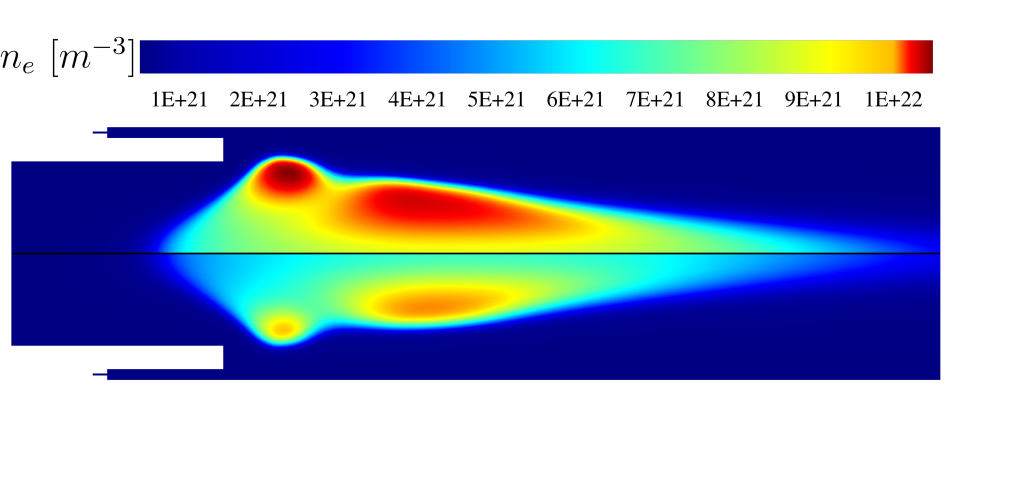}
\caption{Electron number density distribution at \SI{30}{kPa}, and \SI{200}{kW} (with $\eta=50\%$). Top: nitrogen plasma, bottom: air plasma.}
\label{fig:ne_nitrogen_air}
\end{figure}

To quantify the contribution of radiative cooling, the radiative heat loss distribution was integrated over the entire torch
\begin{equation}
Q_{\mathrm{Rad}}=\int \Omega^{\mathrm{R}} d V.
\end{equation}
In ICP simulations, the Joule heating distribution integrated over the torch volume represents the input power (\emph{i.e.,} operating power $\times$ $\eta$) provided to  the plasma via coils:
\begin{equation}
P_{in}=Q_{J}=\int \Omega_{J} d V.
\end{equation}
\cref{fig:Qrad_vs_pr} shows the total radiative heat loss in the torch as a percentage of the input power. The extent of radiative cooling for both gas mixtures increases with pressure, which was expected as radiation losses scale with the number densities of radiating species as discussed previously. Also, nitrogen is found to give a higher radiative cooling as compared to air. For air, Q\textsubscript{Rad} is negligible (less than 2\% of the input power) for pressures below \SI{5}{kPa}. However, at \SI{10}{kPa}, Q\textsubscript{Rad} becomes significant and contributes to $\simeq$ 5\% of the input power, and increases up to 18\% at \SI{101}{kPa}. For nitrogen, Q\textsubscript{Rad} becomes significant at much lower pressure, being 5\% of the input power at \SI{5}{kPa}, and increasing up to $\simeq$ 24\% at \SI{101}{kPa}. The observation that nitrogen plasma has higher radiative cooling is consistent with the fact that nitrogen plasmas have higher concentrations of radiatively active species, especially molecular nitrogen (N\textsubscript{2}) and ionized nitrogen (N\textsubscript{2}\textsuperscript{+}). Nitrogen has strong molecular band emissions, especially in the ultraviolet and visible regions (\emph{i.e.}, second positive system and first negative system), in addition to the strong continuum and line radiation systems of the atomic nitrogen. On the other hand, in air we have only about 79\% nitrogen and 21\% oxygen, whose dissociation/ionization products $\left(\mathrm{O}, \mathrm{O}_2^{+}\text{, and } \mathrm{O}^{+}\right)$ generally radiate less efficiently than their nitrogen counterparts. Also, because radiation often scales like $\propto n_{\mathrm{e}}^2$ \cite{wilbers1991radiative,ogino2013fitting} and depends on excited states, differences in electron density and excitation rates between nitrogen and air plasmas also matter. \cref{fig:ne_nitrogen_air} compares the electron number density distribution in the torch at \SI{30}{kPa} for nitrogen and air plasmas, showing a significantly higher electron concentration in the nitrogen plasma.

\subsubsection{Effect of radiation loss on temperature profiles}
\begin{figure}[!htb]
\centering
\subfloat[][]{\includegraphics[scale=0.2]{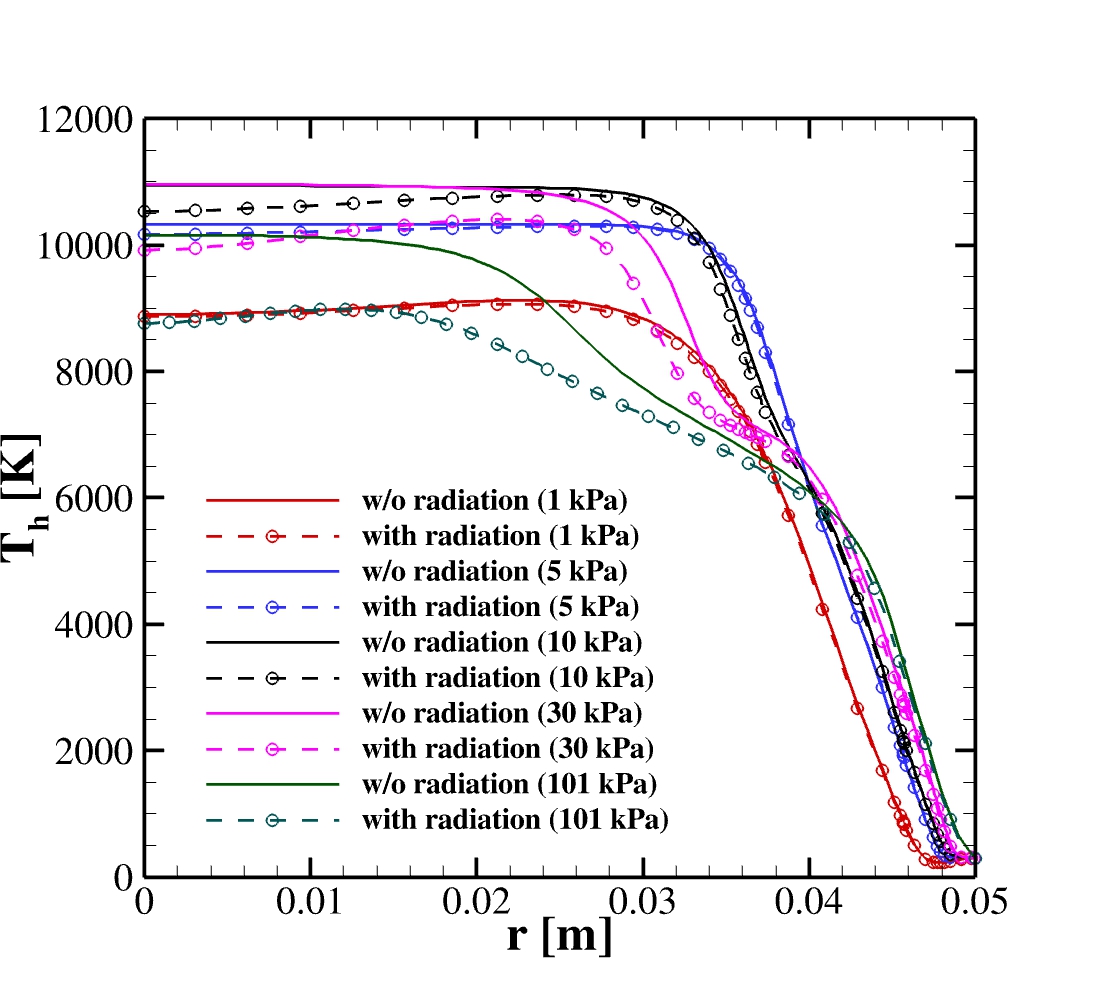}} 
\subfloat[][]{\includegraphics[scale=0.2]{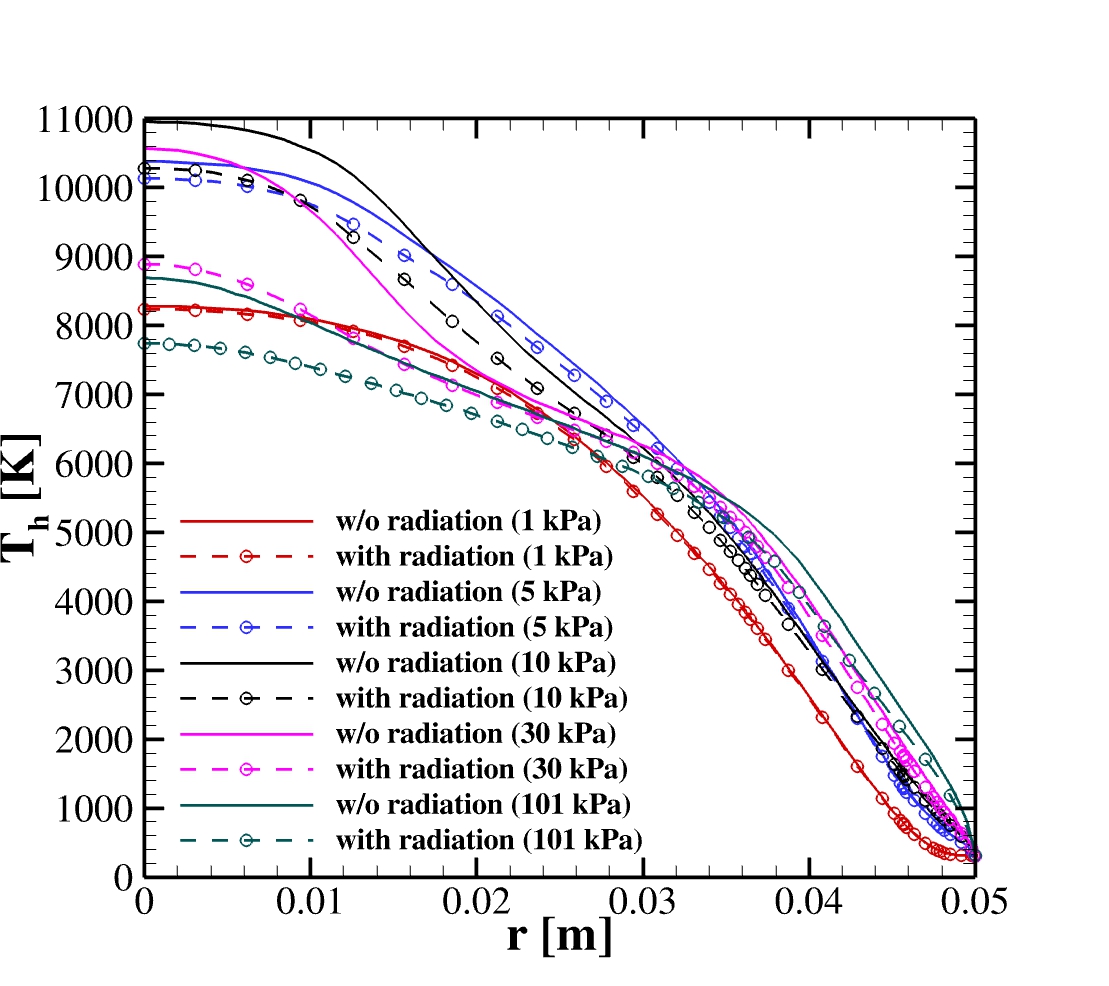}}
\caption{Radial profiles of heavy-species temperature for nitrogen plasma at a fixed operating power of \SI{200}{kW} (with $\eta = 50\%$): (a) $x = \SI{0.15}{m}$ (mid-torch location) and (b) $x = \SI{0.36}{m}$ (nozzle exit).} 
\label{fig:Th_profiles_N2}
\end{figure}

\begin{figure}[!htb]
\centering
\subfloat[][]{\includegraphics[scale=0.2]{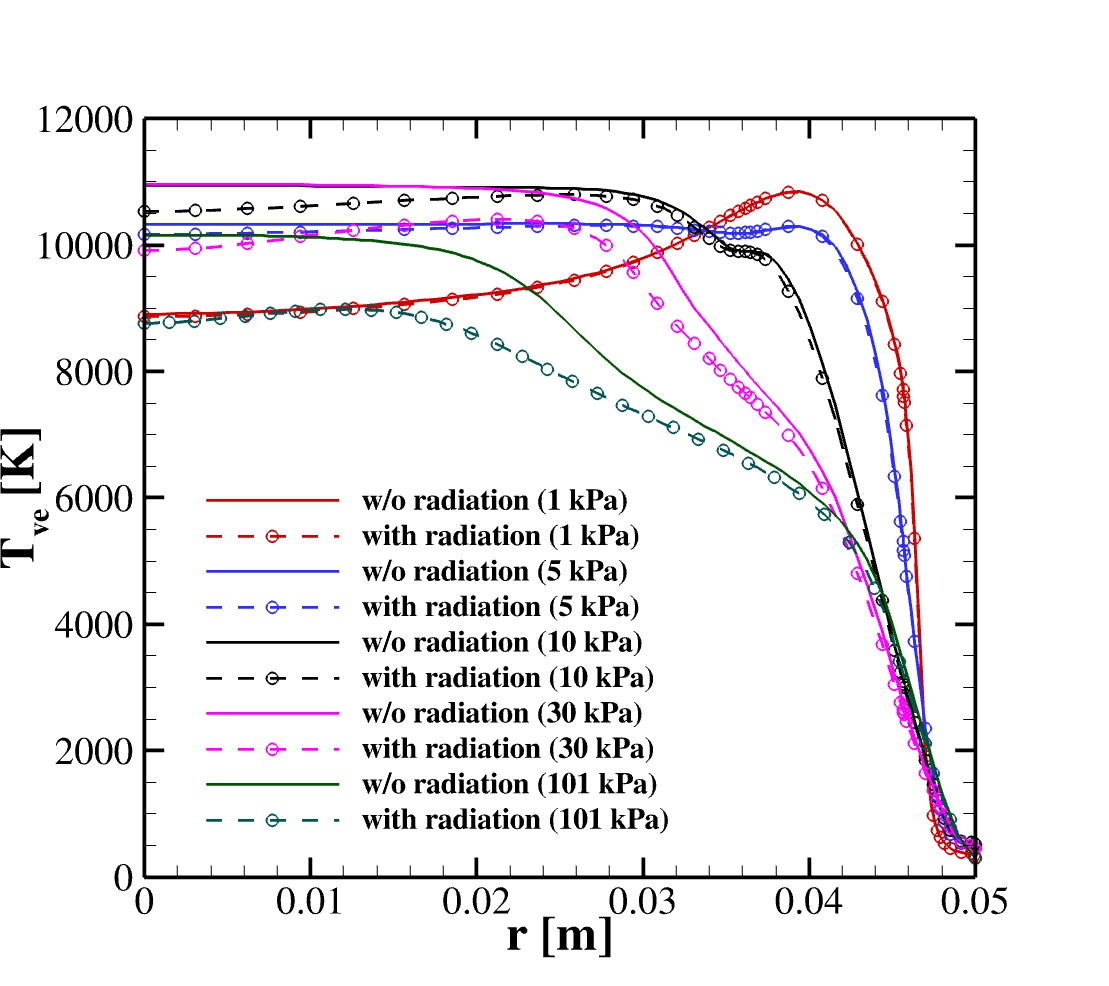}} 
\subfloat[][]{\includegraphics[scale=0.2]{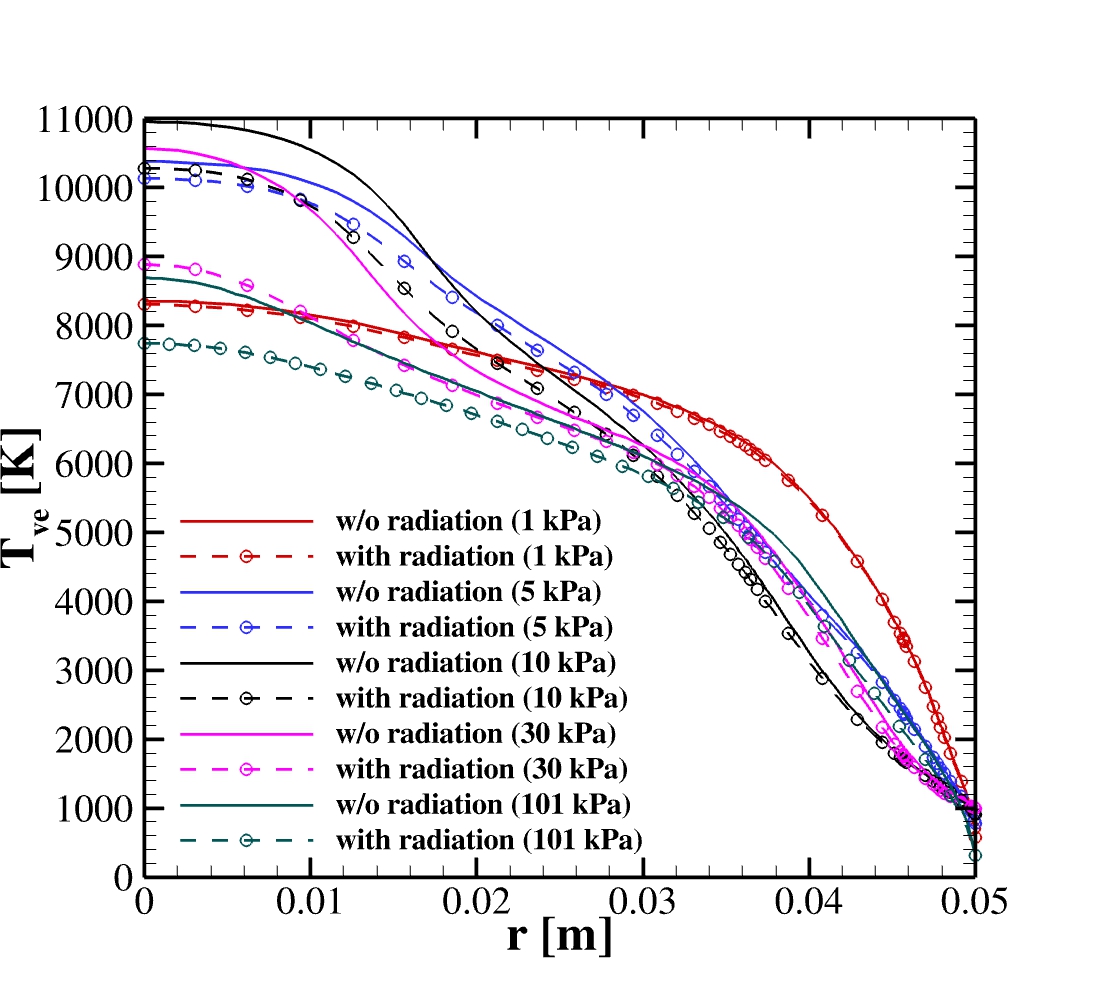}}
\caption{Radial profiles of electro-vibrational temperature for nitrogen plasma at a fixed operating power of \SI{200}{kW} (with $\eta = 50\%$) : (a) $x = \SI{0.15}{m}$ (mid-torch location) and (b) $x = \SI{0.36}{m}$ (nozzle exit).} 
\label{fig:Tev_profiles_N2}
\end{figure}

\begin{figure}[!htb]
\centering
\subfloat[][]{\includegraphics[scale=0.2]{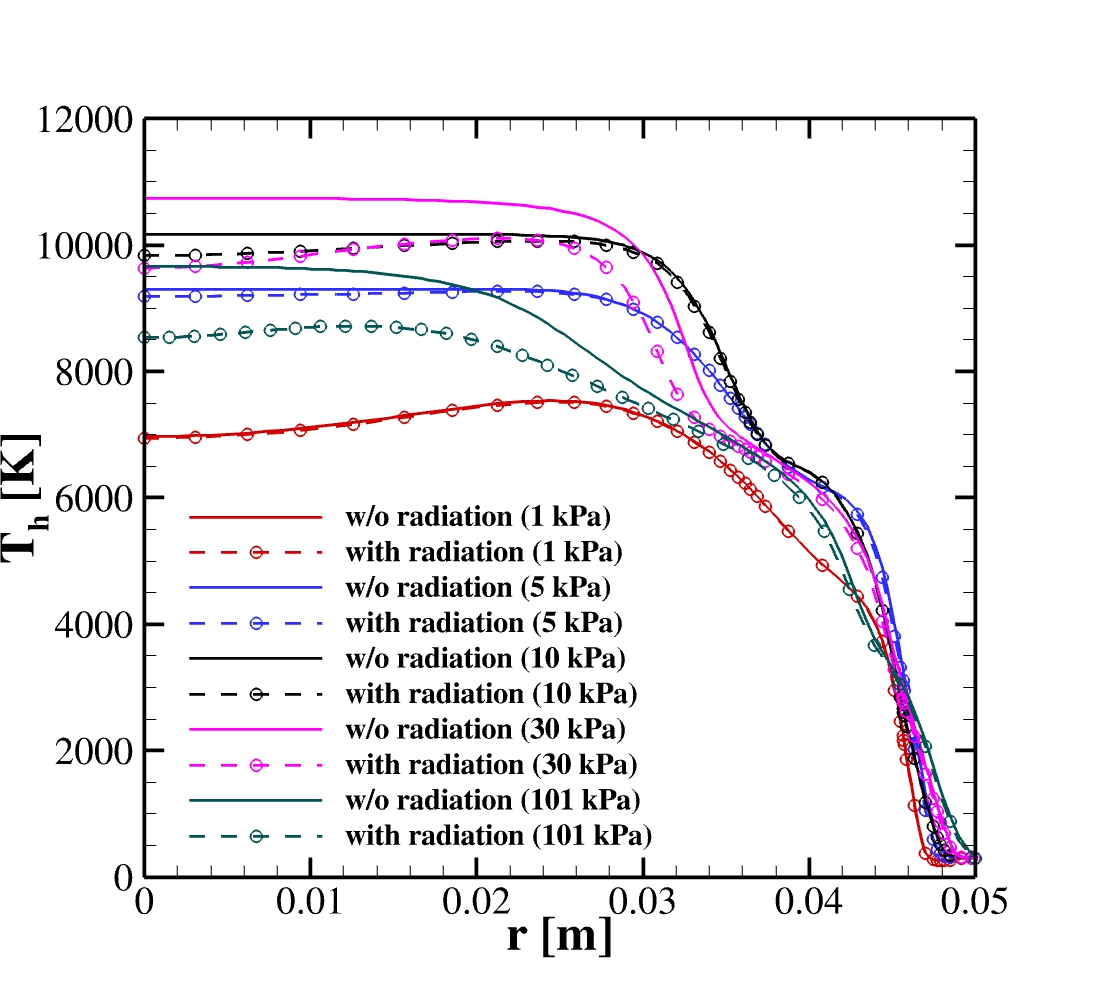}} 
\subfloat[][]{\includegraphics[scale=0.2]{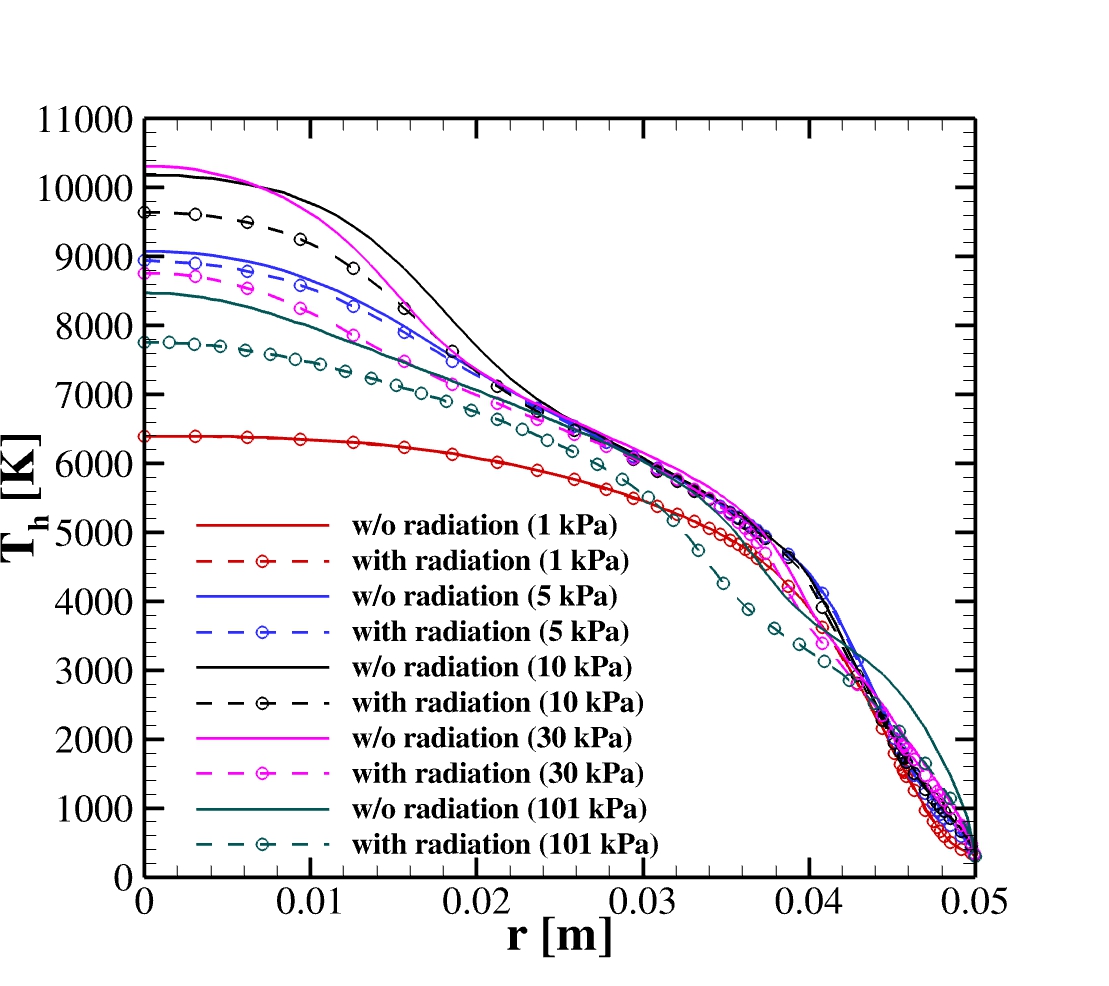}}
\caption{Radial profiles of heavy-species temperature for air plasma at a fixed operating power of \SI{200}{kW} (with $\eta = 50\%$) : (a) $x = \SI{0.15}{m}$ (mid-torch location) and (b) $x = \SI{0.36}{m}$ (nozzle exit).} 
\label{fig:Th_profiles_air}
\end{figure}

\begin{figure}[!htb]
\centering
\subfloat[][]{\includegraphics[scale=0.2]{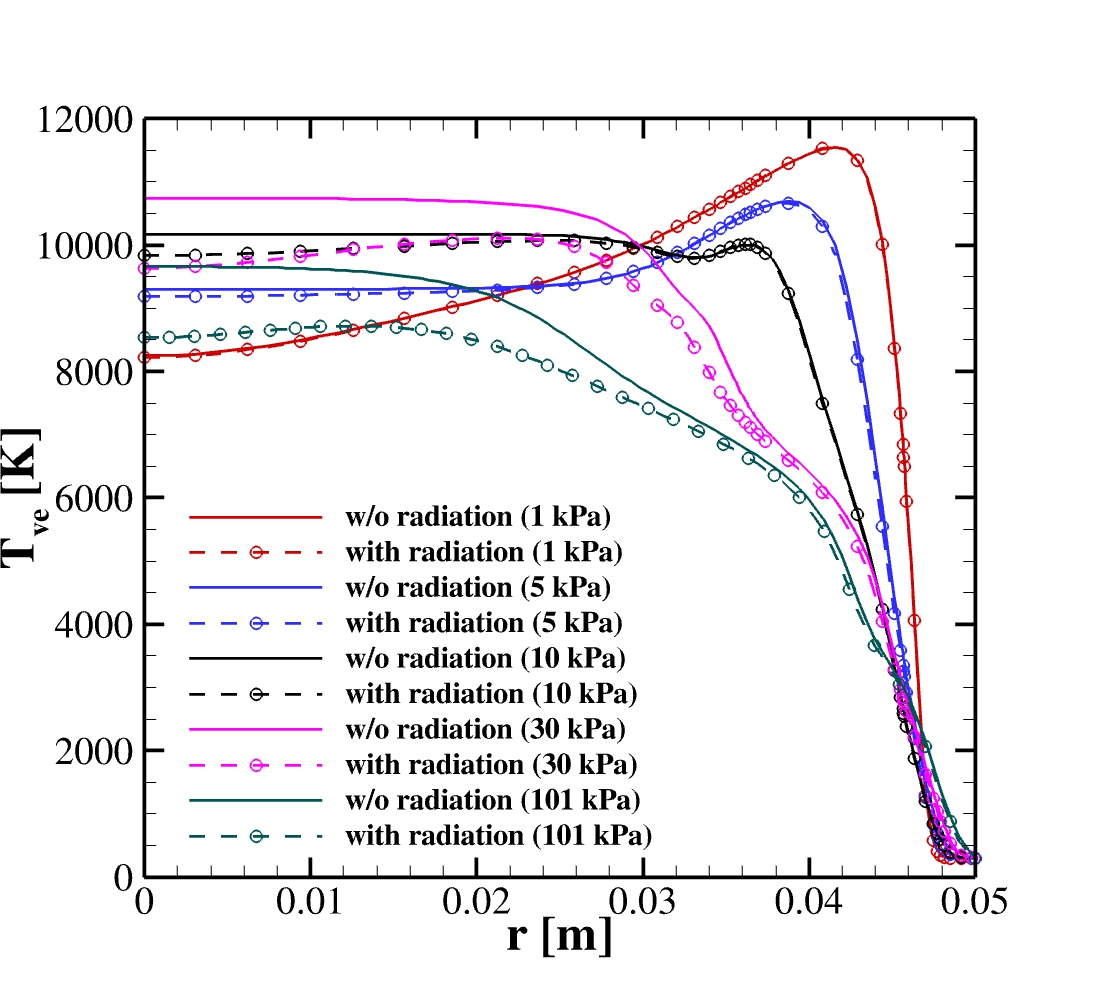}} 
\subfloat[][]{\includegraphics[scale=0.2]{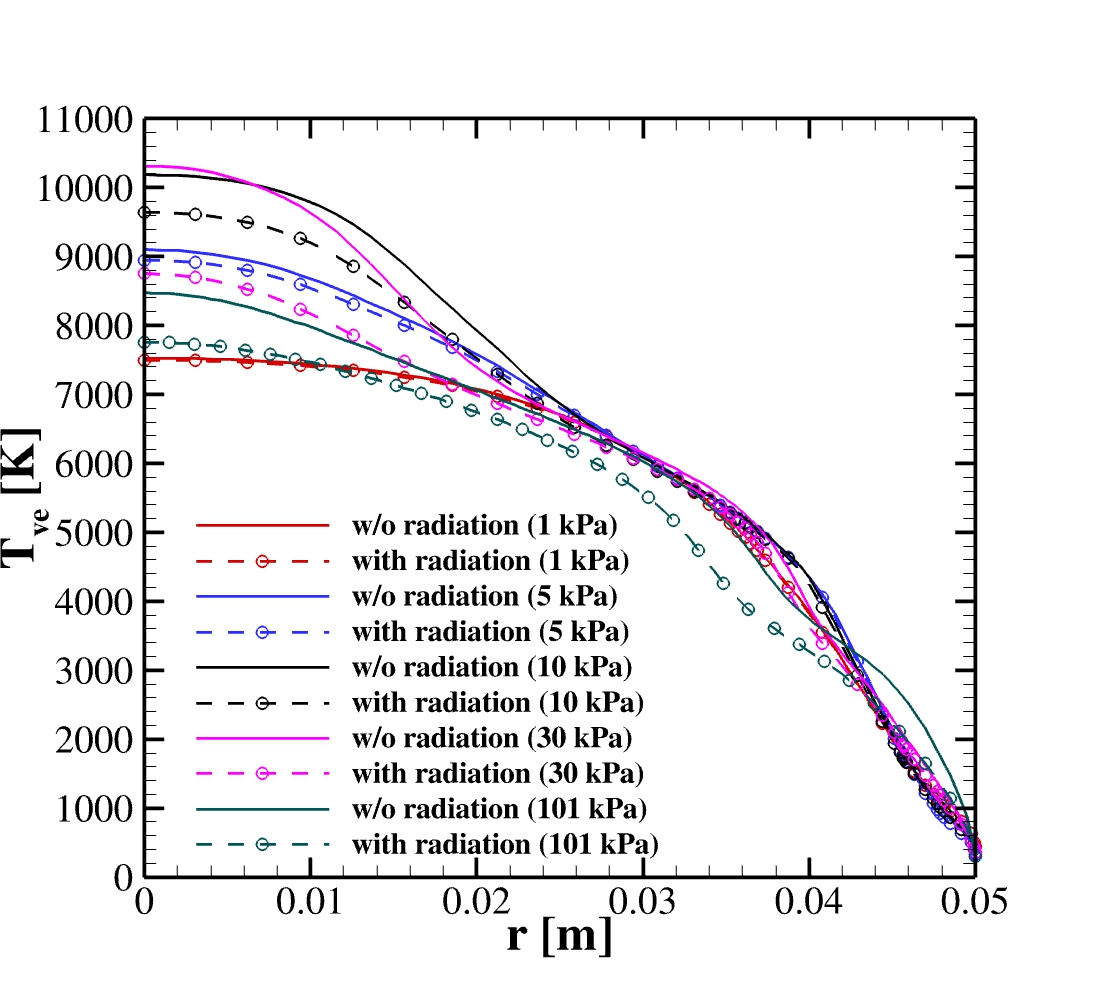}}
\caption{Radial profiles of electro-vibrational temperature for air plasma at a fixed operating power of \SI{200}{kW} (with $\eta = 50\%$) : (a) $x = \SI{0.15}{m}$ (mid-torch location) and (b) $x = \SI{0.36}{m}$ (nozzle exit).} 
\label{fig:Tev_profiles_air}
\end{figure}

\cref{fig:Th_profiles_N2,fig:Tev_profiles_N2,fig:Th_profiles_air,fig:Tev_profiles_air} show the radial profiles of heavy-particle and electro-vibrational temperatures at $x = \SI{0.15}{\meter}$ (mid-torch location) and at the torch outlet for both nitrogen and air plasmas. To avoid confusion in the discussion, we focus on the outlet profiles, as these are more relevant for the characterization of the aerothermal response of TPS samples. Moreover, since the plasma is essentially in thermal equilibrium at the torch exit (\emph{i.e.}, $T_{\mathrm{h}} \simeq T_{\mathrm{ve}}$), there is no need for a separate discussion of the two temperatures. The reported profiles show that temperature drops are significant near the axis. On the other hand, close to the wall, the radiation-coupled profiles collapse onto the uncoupled simulation results for all the pressures. In light of this, the impact of radiation is here quantified in terms of the temperature difference on the axis. The profiles show a negligible effect of radiation at \SI{1}{kPa} and \SI{5}{kPa}. However, beyond \SI{10}{kPa}, radiation losses become significant and lead to temperature drops of around \SI{800}{K} (7.2\%) and \SI{600}{K} (5.8\%) for nitrogen and air plasmas, respectively. At \SI{30}{kPa}, the decrease in temperature due to radiation losses is much higher, and around \SI{1700}{K} (16\%) for nitrogen and \SI{1600}{} (15.3\%) for air. Finally, at 1 atm, the temperature drop is around \SI{900}{K} (10.3\%) and \SI{600}{K} (7.14\%) for nitrogen and air plasmas, respectively. It is interesting to note that, although the temperature drop at the axis at 1 atm is smaller than that at \SI{30}{kPa}, there is a significant temperature drop throughout the radial direction in this case, an indication of an overall higher total radiative power loss.

\subsubsection{Pressure-power mapping of the total radiative power loss}
Since radiative losses in ICPs are governed primarily by temperature and number density of radiating species, they exhibit a strong dependence on the operating pressure, as also demonstrated in the foregoing discussion. This is because the number density of particles scales directly with pressure (\emph{i.e.}, $n = p/ k_{\textsc{b}} T$). In contrast, the dependence of radiation loss on operating power is expected to be comparatively weaker. As a matter of fact, increasing the input power at a fixed pressure typically leads to only a moderate rise in temperature and electron number density due to enhanced ionization. Nevertheless, in the use of ICP facilities, it is common to perform both power and pressure sweeps for the purpose of reproducing a spectrum of flight conditions as broad as possible. Therefore, constructing a two-dimensional pressure–power map of the total radiative power loss may provide valuable insight into how these parameters jointly influence the impact of radiation on the plasma discharge. Such a map not only benefits researchers working with the Plasmatron X facility but also serves as a useful reference for the broader ICP community, enabling informed decisions prior to undertaking computationally intensive radiation-coupled simulations. Consequently, this section investigates how the total radiative power loss in the Plasmatron X facility depends on these two key operational parameters (pressure and power). 

\begin{figure}[!htb]
\centering
\subfloat[][]{\includegraphics[clip,bb=200 57 595 340,width=0.55\textwidth]{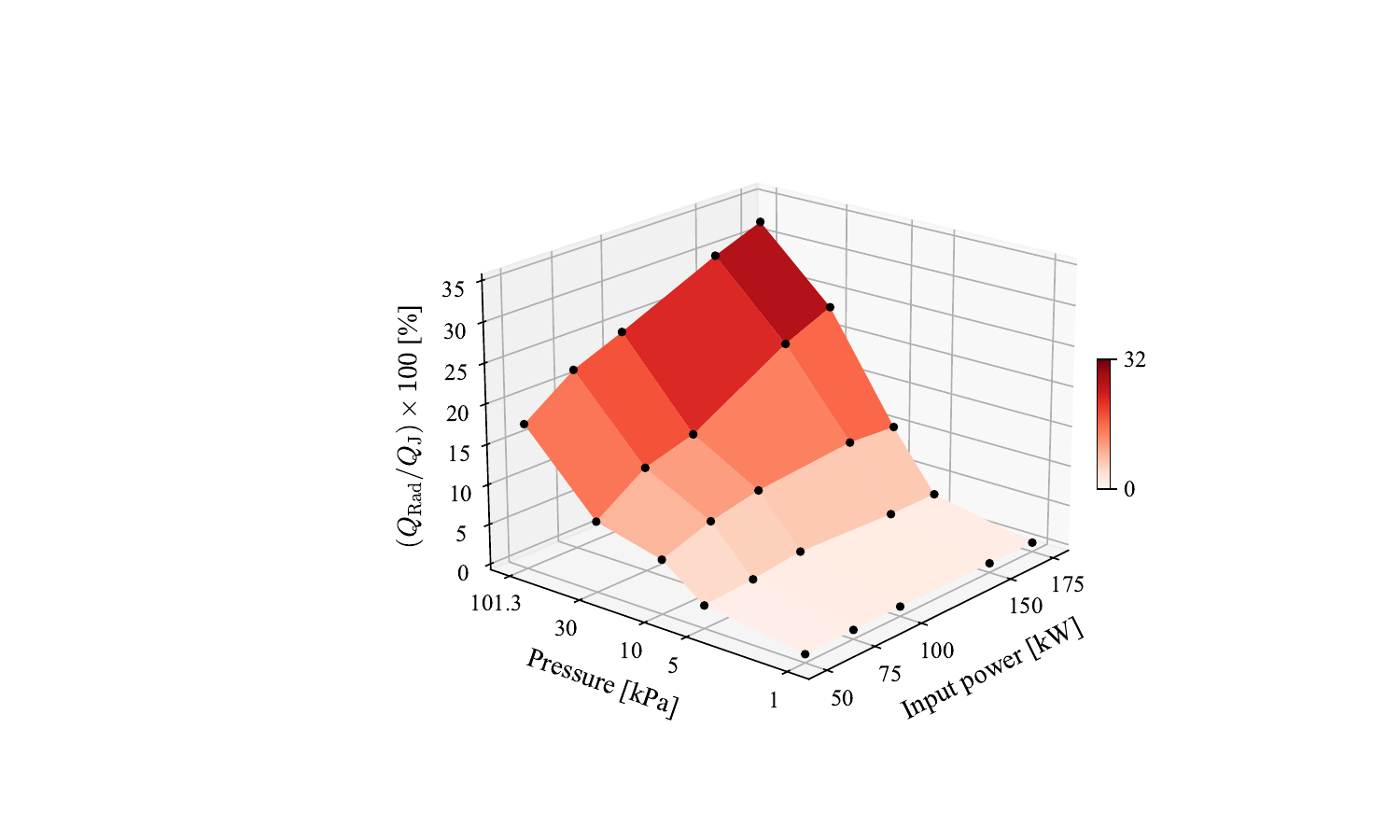}} 
\subfloat[][]{\includegraphics[clip,bb=200 57 595 340,width=0.55\textwidth]{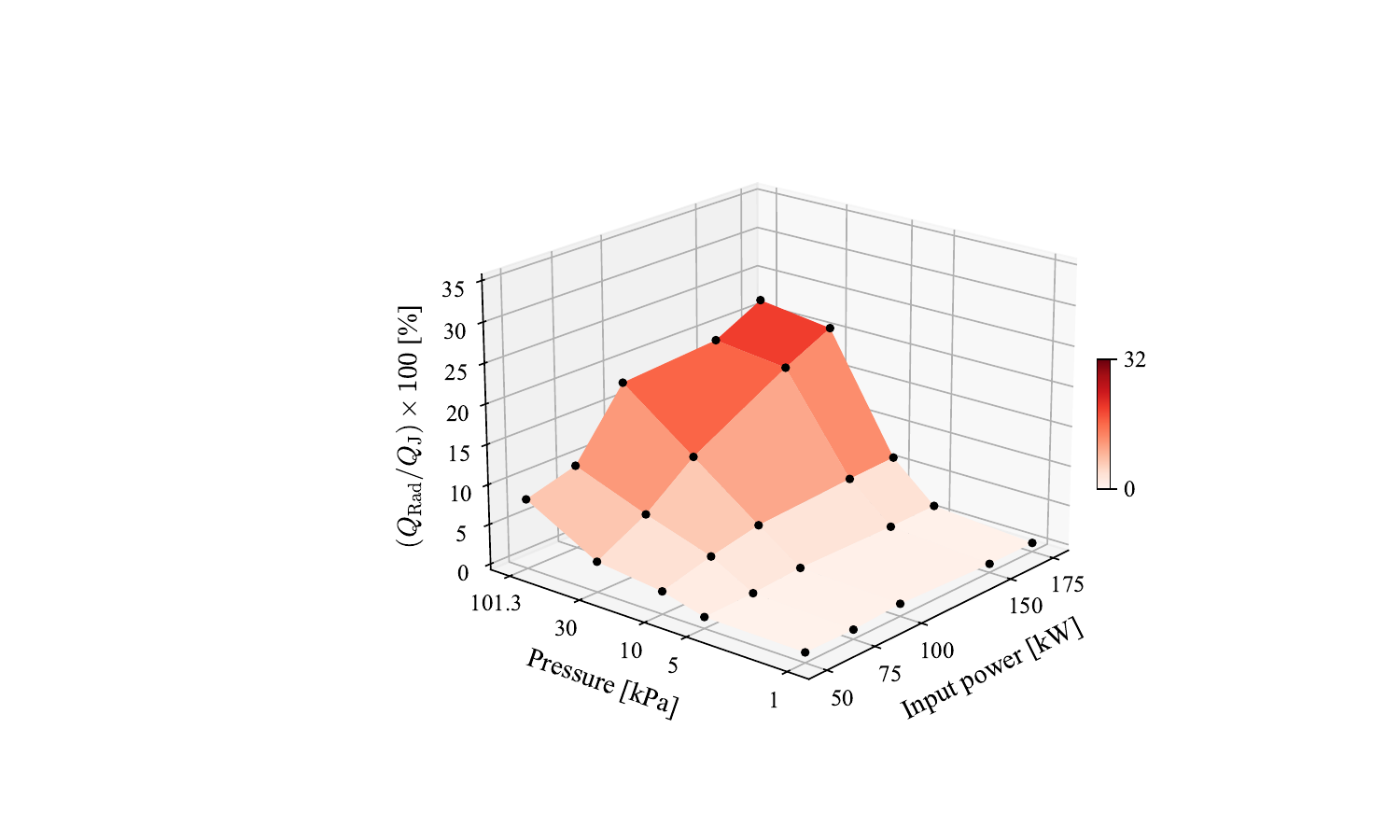}}
\caption{Total radiative heat loss in the ICP torch as a percent of input power : (a) nitrogen, and (b) air plasmas. The black dot implies the actual data point for which the simulation is carried out.} 
\label{fig:2D_Qrad_Map}
\end{figure}

A two-dimensional Cartesian grid of operating conditions has been constructed by considering the following values of pressure and operating power: $p_{\mathrm{a}} \in [1, 5, 10, 30, 101]$ kPa, $P \in [100, 150, 200, 300, 350] $ kW. The efficiency is assumed to be 50\% for all the cases. Hence, the input power to the plasma is half of the operating power, \emph{e.g.}, $P_{\mathrm{in}} \in [50, 75, 100, 150, 175]$ kW. \cref{fig:2D_Qrad_Map} illustrates the two-dimensional pressure–power map of the total radiative power loss, expressed as a percentage of the input power, for both nitrogen and air plasmas. A clear trend emerges for both mixtures: the radiative power loss depends strongly on the operating pressure across the entire range of input powers, whereas its dependence on input power remains relatively weak at lower pressures (\SI{10}{kPa} and below). This behavior aligns well with expectations, since at lower pressures, the particle number density, and consequently, the number of radiating species, is limited, leading to comparatively modest changes in radiation even when the input power is varied. At higher pressures, however, the situation changes significantly. The radiative power loss shows a pronounced increase with increasing input power, reflecting the combined effect of enhanced particle density and high temperatures that promote stronger emission. For nitrogen, the radiative loss rises from approximately 8\% at \SI{50}{kW} to about 23\% at \SI{175}{kW} at an operating pressure of \SI{30}{kPa}, and from 17\% to nearly 32\% over the same power range at \SI{101}{kPa}. A similar but somewhat less pronounced trend is observed for air plasma, where the radiative loss increases from around 3\% to 20\% between 50 and \SI{175}{kW} at \SI{30}{kPa}, and from 8\% to 22\% at \SI{101}{kPa}. The above results clearly demonstrate that, while pressure remains the primary factor governing the magnitude of radiative losses, the influence of power becomes increasingly significant at elevated pressures. The combined pressure–power dependence revealed in this mapping thus provides valuable insight into the radiation behavior of both nitrogen and air plasmas under different operating conditions. 

\begin{figure}[h]
\begin{center}
\includegraphics[width=0.55\textwidth]{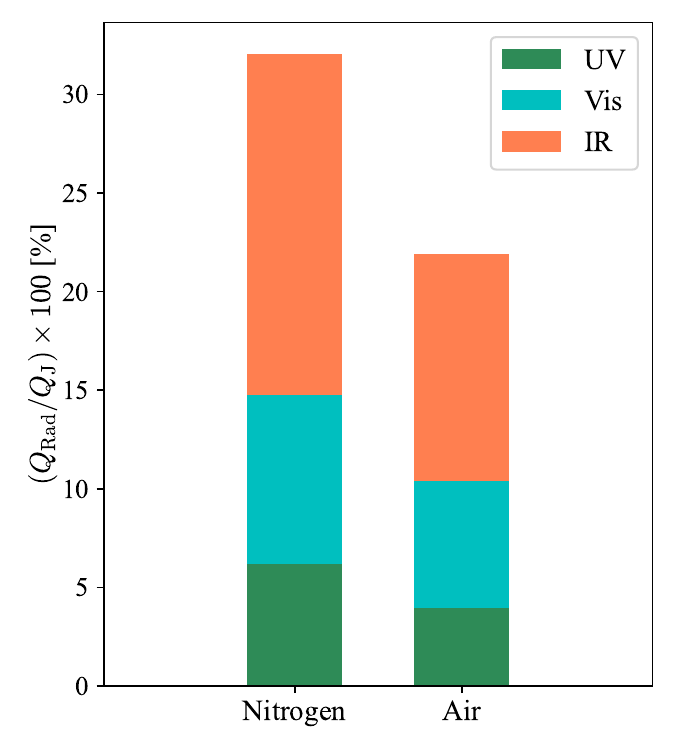}
\caption{Decomposition of the total radiative heat loss in the ICP torch as a percent of input power into the UV, Visible, and IR wavelength ranges. ($p_{\mathrm{a}} = \SI{101.3}{kPa}$, $P = \SI{350}{\kilo\watt}$, $\eta = 50\%$).   }
\label{fig:Qrad_mapping_Decomposition}
\end{center}
\end{figure}

Finally, \cref{fig:Qrad_mapping_Decomposition} shows the wavelength-dependent decomposition of the percentage total radiative power loss at the highest power and pressure condition. It is clear that infrared (IR) radiation contributes most to the total radiation loss for both nitrogen and air. The low influence of the ultraviolet (UV) range implies that the majority of the radiation loss is due to atomic lines and continuum transitions of N and O, rather than the molecular band systems that are dominant in the UV range. This information may provide apriori insight for the design of an experiment using optical diagnostics for an ICP facility to define a measurement wavelength regime.

\subsubsection{Assessment of the plasma optical thickness}\label{sec:optical_thickness}
\begin{figure}[h]
\begin{center}
\includegraphics[scale=0.4, clip, trim=0in 1in 0in 1in]{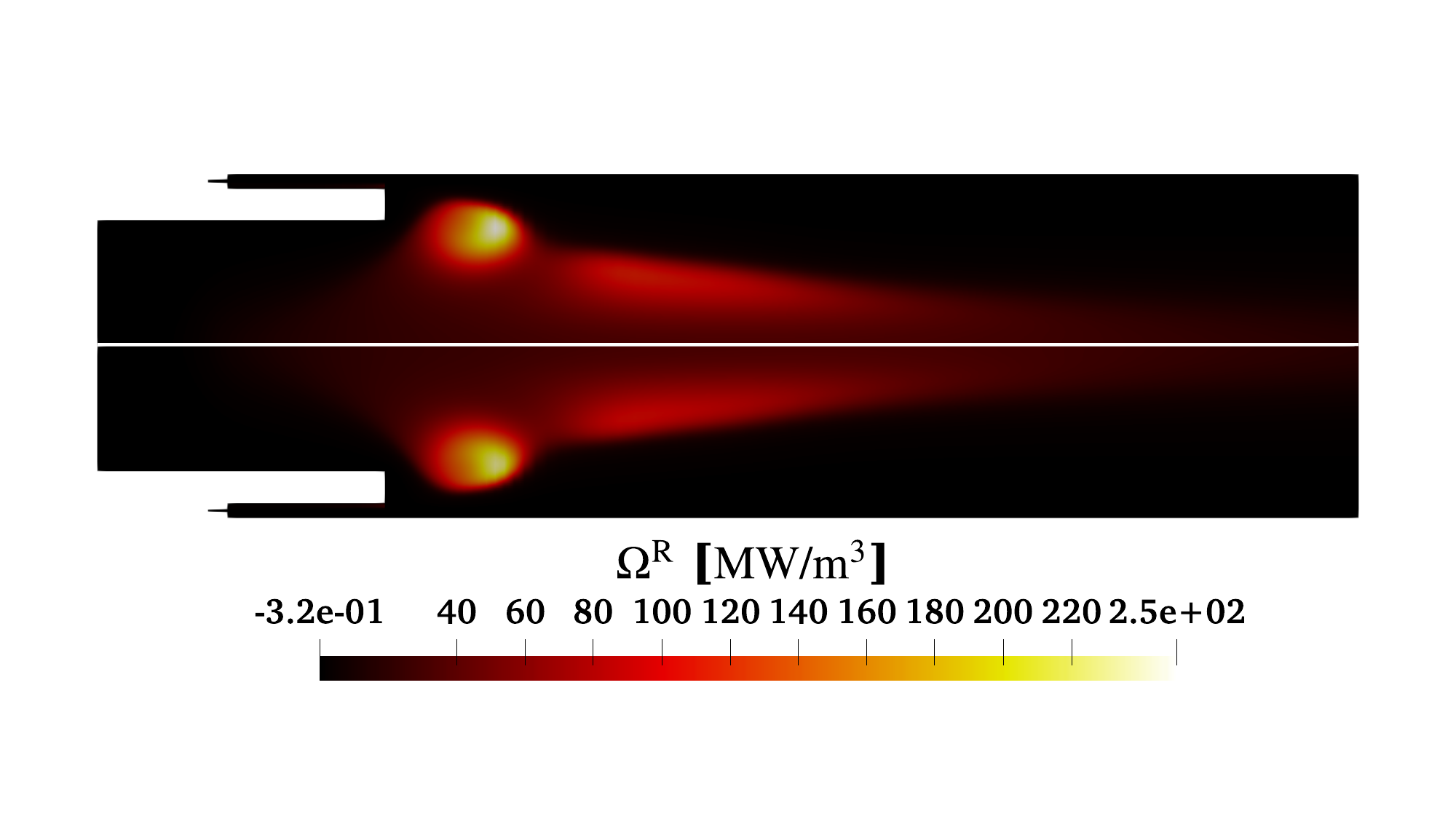}
\caption{Radiative heat loss distribution in the torch for air plasma. Top: without self-absorption (optically thin assumption), bottom: with self-absorption. ($p_{\mathrm{a}} = \SI{101.3}{kPa}$, $P = \SI{350}{\kilo\watt}$, $\eta = 50\%$).   }
\label{fig:optically_thin}
\end{center}
\end{figure}

To assess whether the plasma inside the Plasmatron X torch operates in an optically thin or optically thick regime, a set of radiation-coupled torch simulations was performed for the highest power and pressure conditions (\emph{i.e.}, \SI{101.3}{kPa}, \SI{350}{kW} (with $\eta=50\%$)) for both nitrogen and air plasmas. In these simulations, the self-absorption option in the RTE solver was disabled, thereby neglecting in-plasma reabsorption of emitted radiation and effectively enforcing an optically thin approximation. The resulting solutions were then compared against the baseline radiation-coupled simulations in which self-absorption was accounted for. The highest power–pressure operating point was selected for this analysis because radiative heat losses are maximum under these conditions, making this case the most sensitive for distinguishing between optically thin and optically thick behavior. \cref{fig:optically_thin} compares the spatial distributions of radiative heat loss in the torch for air plasma obtained with and without accounting for self-absorption. The two distributions are essentially identical, indicating that reabsorption has a negligible impact on the radiative field. A similar behavior was observed for nitrogen. Quantitatively, the total radiative power loss as a percentage of input power, expressed as $(Q_{Rad}/Q_J)\times100$, is 22.72\% for air plasma when self-absorption is neglected, compared to 21.89\% when self-absorption is taken into account. For nitrogen, the corresponding values are 32.6\% and 32.05\%, respectively. The small differences between these cases confirm that in-plasma reabsorption of radiation is negligible and that most of the emitted radiation escapes the plasma without significant attenuation. These findings therefore demonstrate that the Plasmatron X torch operates in an optically thin regime and provides a strong basis for future work aimed at developing simplified correlations or reduced-order curve-fit models for the radiative loss term. Such models, derived from detailed numerical simulations and theoretical analyses, could significantly reduce computational cost while retaining the essential physics governing radiation transport, thereby enabling more efficient implementation for high-fidelity simulations of high-enthalpy plasmas.

%% file: Conclusions.tex
\section{Conclusions}\label{sec:conclusion}
This work presented a comprehensive numerical investigation of radiative heat transfer in a high-power inductively coupled plasma wind tunnel using a loosely coupled, non-equilibrium, multi-physics modeling framework. A key contribution of the study is the direct coupling of an NLTE magnetohydrodynamic plasma framework with a spectrally resolved RTE solver, enabling the explicit treatment of emission, absorption, and transport of radiation without relying on optically thin assumptions or empirical radiative loss models. To the authors’ knowledge, this represents the first systematic application of coupled radiation–CFD simulations to an ICP facility.

The developed framework was applied to the \SI{350}{\kilo\watt} Plasmatron X facility at the University of Illinois Urbana-Champaign for both nitrogen and air plasmas over a broad range of operating pressures and powers. The results demonstrate that radiation transport is negligible at low pressures (\emph{i.e.}, $\leq$ 5 kPa), demonstrating the validity of the simplifying assumptions commonly employed in prior ICP modeling efforts under such conditions. However, as the operating pressure is increased, radiative cooling becomes increasingly significant and ultimately constitutes a major energy loss mechanism. At atmospheric pressure, radiation accounts for up to approximately 32\% of the input power for nitrogen and 22\% for air, leading to substantial reductions in plasma core temperatures and pronounced modifications of radial temperature distributions.

A clear distinction between nitrogen and air plasmas was observed, with nitrogen exhibiting consistently higher radiative losses. This behavior was attributed to the higher concentrations of radiatively active species, stronger molecular and atomic emission systems, and elevated electron number densities present in nitrogen plasmas. The spatial distribution of radiative losses was found to be closely correlated with regions of high electro-vibrational temperature and electron concentration, underlying the importance of accurately capturing NLTE effects when modeling radiation in ICP facilities.

To provide practical insight for facility operation and future modeling efforts, pressure–power maps of total radiative loss were constructed. These maps reveal that pressure is the dominant parameter governing radiation losses, while the influence of input power becomes increasingly important at elevated pressures. Such mappings offer valuable guidance for anticipating radiation effects before performing computationally intensive radiation-coupled simulations and for interpreting experimental measurements in high-pressure ICP regimes. Further, spectral decomposition of the radiative losses indicates that infrared emission dominates the total radiative power loss for both nitrogen and air plasmas, with comparatively smaller ultraviolet contributions, implying that atomic bound–bound and continuum radiation is the primary loss mechanism and providing clear guidance for selecting optimal wavelength ranges for optical diagnostics.

Finally, an assessment of plasma optical thickness showed that in-plasma self-absorption has a negligible impact on the radiative field, even at the highest pressure and power conditions examined. This confirms that the Plasmatron X torch operates predominantly in an optically thin regime, despite exhibiting large absolute radiative losses at elevated pressures. This finding reconciles the observed importance of radiation with the limited role of reabsorption in the facility and provides confidence in the applicability of optically thin approximations for selected diagnostic and modeling purposes.

Overall, the present study highlights the critical role of radiation in high-pressure ICP operation and establishes a high-fidelity computational framework capable of capturing the complex interplay between plasma dynamics, electromagnetic fields, and non-equilibrium radiation. The insights gained herein are directly relevant to the design, operation, and interpretation of ICP wind tunnel experiments for hypersonic ground testing. Future work will focus on extending the framework to three-dimensional configurations, incorporating more detailed state-to-state kinetics and non-Boltzmann radiation models, and coupling the facility plasma to material response models to enable end-to-end prediction of aerothermal test environments.